\documentclass[]{article}
\usepackage{geometry}
\usepackage{graphicx}
\usepackage{authblk}
\usepackage[]{natbib}
\usepackage{float}
\usepackage{subcaption}
\usepackage{amsmath}
\usepackage{mathrsfs}
\usepackage{units}
\usepackage[percent]{overpic}
\usepackage{pict2e}
\usepackage{xcolor,varwidth}

\title{Structure and Persistence of Ship Wakes and the Role of Langmuir--Type Circulations}

\author[1]{J. Ryan Somero}
\author[2]{Andre Basovich}
\author[1]{Eric G. Paterson}
\affil[1]{Kevin T. Crofton Department of Aerospace and Ocean Engineering, Virginia Tech, Blacksburg, VA 24061}
\affil[2]{Cortana Corporation, Falls Church, VA 22046}
\date{}                     
\begin{document}

 \maketitle

\begin{abstract}
 Surface ships operating in ocean waves are shown to generate transverse  surface currents which persist long after the initial ship--induced currents  have decayed.  These surface currents are due to the formation of large--scale Langmuir--type circulations and are suggested to be the mechanism for the concentration of surface--active substance into streaks or bands, which are regularly seen in synthetic aperture radar imagery of the ocean surface.  These circulations are generated by the Craik--Leibovich vortex force, which results from interaction of ship--induced current with ambient surface waves. Once generated, the circulations persist due to their large--scale nearly--inviscid nature.  Numerical simulations of the wake behind a surface ship in calm, head, and following seas are presented and show good agreement with radar imagery from at--sea experiments.  It is demonstrated that surface currents do not persist in the absence of surface waves, but can persist for tens of kilometers through the formation of Langmuir--type circulations.  Since the circulations are driven by the cross product of the Stokes drift and the wake vorticity, the structure of the persistent wake is a function of relative heading of the ship with respect to the direction of surface--wave propagation.
\end{abstract}


\section{Introduction}
 The study of ship wakes presents a field that has had great theoretical and applied interest for over a century, starting with the description of the Kelvin wake \citep{tho87}.  The theoretical interest is still significant since not all mechanisms related to ship wakes are understood, particularly the long persistence of ship--induced perturbations.  There is a high degree of applied interest in the matter, as ship wakes can be observed in synthetic aperture radar (SAR) imagery \citep{mun87, ree02, bru11} and optical imagery \citep{mun87}.

There are several distinct processes generated by a self--propelled ship: surface waves generated by the ship hull (Kelvin wake); propeller and hull--induced near--surface flow and turbulence; and bubbles produced by the propellers and breaking bow waves \citep{pel84}.  These features are well documented in field \citep{pel93} and towing--tank \citep{pel92,swe87} experiments.  Most of these features decay several ship lengths behind the vessel \citep{swe87}, which is also shown by simulations presented in this paper.  Nevertheless, it is well known that some features of ship wakes can be observed well after the passage of the ship, in some cases tens of kilometers behind the vessel.  This paper addresses the physical mechanism for the formation and resulting structure of such long--lasting ship wakes, and describes their qualitative and some quantitative features.

A distinction should be made between different stages in the development of a ship wake.  Commonly, a near and far wake are described \citep{pat96, ree02, fuj16}.  The near wake consists of perturbations generated in the vicinity of the ship by propellers, appendages, and the hull.  The far wake is the region of the wake where these perturbations evolve and decay behind the ship.  The physical mechanism of persistence of the ship wake beyond this far-field region is not currently explained.  In this paper, we call the part of the wake that exists long after the initial perturbations generated by the ship dissipate, \textit{the persistent wake}.

In SAR images of the sea surface, ship wakes produce one or several streaks along the trajectory of the ship.  The brightness of the streaks, in comparison to the background, depends on the amplitude of the short surface waves, which are responsible for reflection of the radar waves.  Dark and bright streaks correspond to lower and higher amplitude short surface waves, respectively.  The amplitude of the short surface waves is affected mainly by surface currents, near--surface turbulence, and films of surface--active substance (SAS) which alter the elasticity of the ocean surface.  Surface currents can produce some effect on the wave amplitude in the near-- and far--wake regions \citep{fuj16}.   However, since the velocity of surface currents generated in the ship wake is relatively low, the turbulence and SAS play more important roles in the formation of surface images.  In general, the far wake, where initial turbulent perturbations still exist, is normally seen in the image as a dark streak called the centerline wake, which is formed due to damping of the short surface waves by turbulence. 

\begin{figure}
	\centering
    \begin{overpic}[width=\textwidth]{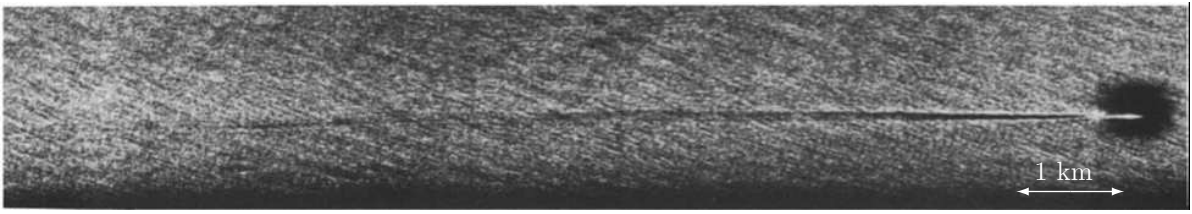}
    \linethickness{0.5 pt}
    \put (90,1.5) {\color{white}\vector(-1,0){4.54}}
     \put (90,1.5) {\color{white}\vector(1,0){4.54}}
     \put (87,2.5) {\color{white} {1 km}}
     \end{overpic}
    \caption{SAR image of USS Chandler (DDG-996) in head seas with a persistent centerline wake \citep{mil93}.}
    \label{fig:HeadSeaSAR}
\end{figure}

\begin{figure}
	\centering
    \begin{overpic}[width=\textwidth]{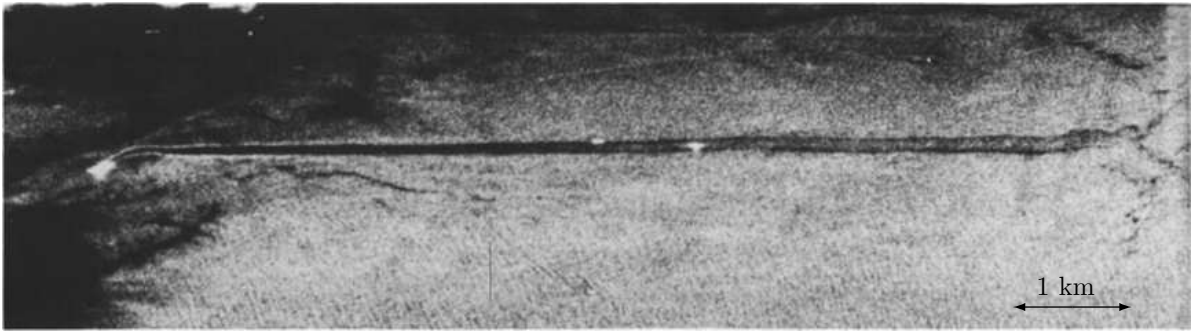}
     \linethickness{0.5 pt}
     \put (90,2) {\vector(-1,0){5}}
     \put (90,2) {\vector(1,0){5}}
     \put (87,3) {1 km}
   \end{overpic}
   \caption{SAR image of USS Chandler (DDG-996) in following seas with a ``railroad--track" persistent wake \citep{mil93}.}
    \label{fig:FollowingSeaSAR}
\end{figure}

Further away from the ship, and depending on environmental conditions, there are two major types of  persistent wakes that could be observed: a ``centerline" wake or a ``railroad--track" wake.  Examples of such ship wakes are presented in figures \ref{fig:HeadSeaSAR} and \ref{fig:FollowingSeaSAR}, which are SAR images from \cite{mil93}.  The centerline persistent wake appears as a dark streak, while the railroad--track persistent wake appears as a bright streak along the centerline with two dark streaks around it (``the railroad tracks").  Under conditions of following seas, it can be seen in the SAR image presented in figure \ref{fig:FollowingSeaSAR}, that in the far wake, the centerline wake transforms into the railroad--track type of persistent wake.  {\textit{In situ}} measurements at the sea surface have demonstrated that the streaks in the persistent wake are caused by the redistribution of SAS films, which dampen the short surface waves reducing their amplitude \citep{pel91, mil93}, and thereby locally change the brightness of the surface imagery.


Redistribution and alignment of SAS into streaks suggests the presence of surface currents with a velocity component transverse to the wake.  \cite{erm14} detected the presence of twin symmetric, persistent circulations in the wake of a ship using an acoustic Doppler current profiler (ADCP).   In the particular case studied by \cite{erm14}, these circulations produced diverging flows (upwelling region) at the centerline on the surface, and downwelling regions at the edges of the wake.  It was concluded that the circulations were responsible for the formation of the streaks with either higher or lower concentration of SAS.  The mechanism for the formation of such circulations, however, was not identified.

There are several hypotheses considered by different authors regarding the nature of the circulations in the region of the persistent wake.  The more frequently suggested mechanisms for the generation of these circulations until now have been: swirl generated by propellers \citep{fuj16}; hull--induced transverse currents; and flow generated by rising bubbles \citep{pel93,erm14}.  However, these mechanisms cannot explain the observed persistent--wake features.  The swirl produced by the propellers decays relatively quickly as demonstrated in this paper by direct simulations, and thus cannot induce the long--lasting surface currents that can cause the persistent wake.  Measurements of the hull--induced transverse currents have not shown persistent circulatory motions that would redistribute surfactants to the wake edges.  For example, towing--tank experiments reported by \cite{swe87} show the formation of a pair of secondary inward rotating vortices with outward rotating propellers that are speculated to be shed from the hull or induced by the propeller swirl, but the same secondary vortex pair is not observed with inward rotating propellers.  Computational simulations of the U.S. Navy ship model 5415 with outward rotating propellers, relative to surface centerline, by \cite{hym98} and \cite{ste01} also show the formation of secondary induced circulatory motions that rotate inward in agreement with the experimental findings of \cite{swe87}.  In addition, the circulations that can be attributed to the redistribution of surfactants to the wake's outer edges must be outward rotating.  Lastly, flows generated by rising bubbles in the wake of the ship are even less studied, and descriptions of the mechanism of their generation are more the result of speculation than of experimental measurement.  The magnitude of bubble--induced fluid velocity, in the case of a reasonable concentration of bubbles, can be expected to be extremely small.

The mechanism of the generation of large circulations (relative to the scale of the ship beam) in a ship wake was first suggested and described by \cite{bas11}.  It was shown that such circulations are produced by the interaction of a near--surface current with ambient surface waves.  The circulations are generated by a vortex force imposed on the ship--generated non--uniform currents by the surface waves \citep{cra76}.  These circulations are similar in nature to Langmuir circulations \citep{lan38, cra76, gar76,sul10,bas14}.  Langmuir circulations (LC) are formed near the sea surface as a result of an instability of the wind--driven surface current interacting with the surface waves.  The non--uniform structure of the current in that case is due to the instability of the developing current.  In contrast, the circulations described by \cite{bas11} are generated around a non--uniform current introduced by an external source.  In the present paper, such circulations are referred to as Langmuir--type circulations (LTC).  Clearly, LTC can be formed as a result of the interaction of ship--generated currents with ambient surface waves.  The characteristic scales of the resulting circulations are much larger than the transverse scale of the current itself \citep{bas11}.  The magnitude and sign of the vortex force imposed on the current induced by the ship depend on the cosine of the angle between the directions of the current velocity and the propagation of the surface waves \citep{cra76}.  Therefore, the structure of the persistent wake will be different for different directions of the ship course relative to the ambient surface wave field.  In the paper by \cite{bas11}, generation and development of LTC were studied using an idealized model of low eddy viscosity, and the effect of viscosity on both the initial current and LTC was neglected.  Three types of currents were considered: a near--surface jet, a near--surface shear current and an underwater jet.  It was demonstrated that the intensity of LTC can be significant under realistic assumptions regarding the magnitude of the initial current and amplitude and wavelength of the ambient surface waves.

In reality, the structure of the current produced by a self--propelled ship is more complex than a simple jet, and the current evolves and decays due to the presence of the strong ship--generated turbulence.  At a minimum, the ship--induced current consists of a propeller--generated thrust current directed away from the ship and a drag current with flow in the direction of the ship motion (e.g., \cite{pao73, has80}).   It is shown in this paper that, in general, there are four circulations produced due to the interaction of the complex ship--induced currents with surface waves.  The resulting system of circulations is unlike the case of a pure jet where only two circulations are generated \citep{bas11}.  The velocity of the initial ship--induced current system decreases with time (and distance from the ship) due to the effect of ship--generated turbulence on the current.  Thus, the development of LTC due to interaction of the current with the ambient surface waves takes place only for a limited time interval.  During the time that the LTC develops, the turbulence generated by the ship decays and eddy viscosity decreases.  Therefore, LTC can persist for a long time (and distance) after the passage of the ship.   All these features of the persistent wake are analyzed in the present paper.  It is demonstrated that the structure of the persistent ship wake depends not only on the current produced by the ship, but also on the surface--wave amplitude and direction relative to the course of the ship.  Under some conditions, the persistent wake may not develop at all.  In short, the persistent wake results from the generation of a secondary flow in the form of LTC due to interaction of a ship--generated thrust-- and drag--current system with the ambient surface wave field.

The paper is structured as follows.  This section presents the statement of the problem.  The next section presents the modeling approach and a description of the numerical algorithms employed in the simulations.  Then the structure of the ship--generated flow field, or system of thrust and drag currents, is described for a particular ship configured to model the naval surface combatant used in the at--sea experimental study of \cite{pel93}.  This ship--generated current system represents the near wake, which is used as the input for CFD simulations of the development of the wake.  Some general features of the current system are also discussed.   Simulations of the wake evolution in the absence of an ambient surface wave field are described.  Then, simulation results, including LTC, for the two general cases of head and following seas are presented.  Existence of the centerline and railroad--track persistent wakes are explained.  The effect of LTC on the redistribution of the SAS film at the surface is discussed, and a comparison of theoretical results with available experimental data is provided.  Results and further required work are also discussed. 

\section{Model formulation}
The formulation of the numerical modeling is based on the goal of simulating wakes that are generated by ships operating in realistic oceanic conditions.  This approach stands in contrast to much of the previous work, which has focused on simple towed or self--propelled vehicle geometries at model--scale Reynolds numbers under highly--stratified laboratory conditions \citep{has80, che99, gou01, dom02, dia05, bru10}.  We employ an extensible computational framework that can include initial conditions specified from geometry--resolving simulations or theoretical models; that can incorporate many types of oceanic forcing; and that is capable of spanning the length and time scales of interest.  While our approach is general, we will limit the discussion to the models, methods, and parameters used for the simulations presented herein.  
To begin, we assume that a ship with length $\textit{L} = \unit[\mathcal{O}(100)]{m}$ is traveling with velocity $U_0 = \unit[\mathcal{O}(10)]{m/s}$ in an Earth-fixed frame of reference and generates a persistent wake with length of $\mathscr{L} = \unit[\mathcal{O}(10)]{km}$.  The Cartesian coordinate system $x_i=(x_1,x_2,x_3)=(x,y,z)$ is aligned such that $x$ is the downstream direction, $y$ is the cross--wake direction, and $z$ is the upward direction with its origin at the ocean surface.  For seawater with a kinematic viscosity of $\nu=\unit[10^{-6}]{m^2/s}$, the corresponding Reynolds and Froude numbers are $Re=U_0L/\nu = \mathcal{O}(10^{9})$ and $Fr=U_0/\sqrt{gL} = \mathcal{O}(0.3)$.  

\subsection{Governing equations}
The governing equations are described by the conservation of mass and linear momentum for an unsteady incompressible flow.  The model is formulated to include oceanic forcing from  current--wave interaction through the Craik--Leibovich (CL) vortex force, originally derived based on time averaging the vorticity transport equation \citep{cra76} and later by applying the Generalized Lagrangian Mean \citep{and87} to the conservation of mass and momentum equations.  This time averaging is applied assuming that the characteristic time scales are long relative to the average wave period, but short relative to the time of formation of the ship induced current.  This requirement suggests that application of the CL vortex force is only appropriate once the axial velocity gradients of the ship induced current have become small, delaying the time and distance behind the ship for where the influence of the force can be observed.  This requirement also suggests that it would not be appropriate to include the vortex force in near field simulations of the ship hydrodynamics where the initial wake field is under formation.

Here we assume that contributions due to Coriolis acceleration and Ekman forcing are small in comparison to the Craik-Leibovich vortex force.  \cite{jon15} demonstrated that in low wind conditions, Ekman forcing is negligible, while in conditions of moderate sea state, Ekman forcing plays a secondary role to Langmuir--type circulations.  Contributions by Coriolis acceleration are considered small here as the characteristic width of the wake is O(100m).

While our simulation toolkit accommodates large--eddy simulation, the length and time scales of interest dictate that unsteady Reynolds averaging be used to model the effects of turbulence.  Decomposing the instantaneous velocity field into moving mean $U_i$ and fluctuating $u_i^\prime$ components, i.e., $u_i = U_i + u_i^\prime$, and taking the time--average gives the following mathematical model

\begin{align}
\frac{\partial U_j}{\partial x_j} & = 0 \label{eq:contRAS},\\
\frac{\partial U_i}{\partial t} + \frac{\partial (U_i U_j)}{\partial x_j}& = -\frac{1}{\rho_0}\frac{\partial \hat{p}}{\partial x_i} + \nu\frac{\partial^2 U_i}{\partial x_j \partial x_j} + \frac{\partial}{\partial x_j} \overline{u^\prime_i u^\prime_j}+\frac{\Delta\rho}{\rho_0}g_j\delta_{ij} + \epsilon_{ijk} U^{St}_j \omega_k \label{eq:momRAS},
\end{align}
where $\hat{p}=\overline{p}-\rho_0 g z$ is the mean piezometric pressure, $\overline{p}$ is the mean total pressure, $g_j$ is the gravitational vector (which is assumed to be pointing downward in the negative $z$ direction), $\nu$ is the kinematic viscosity, $\delta_{ij}$ is the kronecker delta, $\omega_k = \epsilon_{ijk} \frac{\partial U_i}{\partial x_j}$ is the mean fluid vorticity, $\epsilon_{ijk}$ is the permutation tensor for defining the cross--product in tensor notation, and $U^{St}_j$ is vector of the Stokes--drift velocity generated by ocean waves.  

In the context of the unsteady Reynolds-averaged Navier-Stokes (URANS) equations, we are forced to assume that the resolved unsteadiness in the simulation is sufficiently removed from the period over which mean variables are averaged.  Flows considered here have Reynolds numbers on the order of $10^9$, such that there is a considerable variation between viscous scales and integral scales with a time scale ratio($\nicefrac{f_L}{f_{\nu}}$) of O($10^4$).  Characteristic Kolmorgorov length scales are on the order of $10^{-5}m$, while characteristic integral length scales are on the order of $1m$.  To compute the Reynolds stresses $\overline{u^\prime_i u^\prime_j}$, we use the standard eddy--viscosity model $-\overline{u^\prime_i u^\prime_j} = 2\nu_t S_{ij} - \frac{2}{3}k\delta_{ij}$, which is a linear closure relating the Reynolds stresses to the mean rate of strain $S_{ij} = \frac{1}{2}\left(\frac{\partial U_i}{\partial x_j} + \frac{\partial U_j}{\partial x_i}\right)$.   



The last term in equation (\ref{eq:momRAS}) is the Craik--Leibovich vortex force (normalized by water density) that results from current--wave interaction. It is a function the Stokes--drift velocity produced by the surface waves.  For the work here, we consider monochromatic surface waves in deep water.  In this case, a magnitude of the Stokes--drift velocity can be written as
\begin{equation}
U^{St} = \left(2\pi\frac{a_s}{\lambda}\right)^2\sqrt{\frac{g\lambda}{2\pi}}\exp\left(-4\pi\frac{z}{\lambda}\right),
\end{equation}
where $a_s$ is the wave amplitude, $\lambda$ is the wave length. The Stoke--drift velocity vector is horizontal and has only two components: $U^{St}_x = U^{St}\cos{\theta}$ and $U^{St}_y = U^{St}\sin{\theta}$, where $\theta$ is angle between the Stokes--drift velocity and the direction of the ship--produced current ($x$ axis).

In order to better understand generation of LTC in a ship wake and the results of numerical simulations presented in this paper, it is instructional to look at the components of the vortex force. Here we consider the vortex force for a case of two--dimensional current, which is a basic case for all further analysis, with velocity $U(y,z)$ in $x$ direction that is a function of $y$ and $z$ coordinates.  For this case it is easy to see that components of the vortex force corresponding to the last term in equation (\ref{eq:momRAS}) are    
\begin{align}
F_{x} = -\rho_0 U^{St}_y\frac{\partial{U}}{\partial{y}},\quad F_{y} = \rho_0 U^{St}_x\frac{\partial{U}}{\partial{y}},\quad   F_{z} = \rho_0 U^{St}_x\frac{\partial{U}}{\partial{z}}.
\end{align} 

Horizontal and vertical components of the vortex force $F_{y}$ and $F_{z}$ transverse to the current velocity are responsible for generation of the circulations. The component $F_{x}$ will affect the velocity of the initial current. The transverse components $F_{y}$ and $F_{z}$ are proportional to $x$ component of the Stokes--drift velocity and the gradient of the current velocity in the direction of that component.  The presence of a non--uniform gradient in the initial current, e.g., from a practical ship wake, produces a non--uniform vortex force field which in turn will generate the circulations.  The value of the angle $\theta$ between the ship trajectory and the wave direction is of critical importance.  Clearly, the magnitude of $F_{y}$ and $F_{z}$ and corresponding rate of growth of the LTC will be maximum when the Stokes--drift velocity is parallel or anti-parallel to the direction of the current; $\theta=0$ or $\theta=180^\circ$. In the case of a $180^\circ$ change in ship course, the vortex force and resultant circulations will change sign and direction of rotation, respectively. For waves orthogonal to the ship track ($\theta=90^\circ$), the transverse components of the vortex force are zero, and we could therefore expect the current--wave interaction to vanish.  At the same time $F_{x}$ component of the vortex force disappears for the cases of $\theta=0$ or $\theta=180^{0}$ and its effect on the initial current is negligible. Therefore, in this study we focus on the two cases of head seas and following seas.  

\subsection{Computational approach}
While the governing equations are fully unsteady and three dimensional, simulation over both multiple hours of time and sufficiently large domains, while resolving centimeter--scale gradients, is not tractable with current--generation supercomputers.  Therefore, we use a ``two--dimensions plus time" (2D+t) approach, where the three--dimensional problem is reduced to two spatial dimensions, under the assumptions that the wake is slowly evolving in the axial direction, and that characteristic scale of the change of wake parameters (such as velocity and turbulent kinetic energy) in the direction of the wake is much larger than the characteristic width of the wake.  In addition, application of the CL vortex force requires that the time scale for formation of the current is long compared with the period of the surface waves.  These conditions are satisfied if the ship speed is much larger than both the axial defect velocity and the transverse velocity perturbations, which is always the case for a fast--moving ship.

The computational mesh is designed and the boundary conditions are selected to force axial gradients to zero, and therefore facilitate 2D+t simulations in an otherwise fully--general CFD solver.   This is accomplished by using periodic boundary conditions in the axial $x$ direction, and constructing a mesh which is only one--cell thick in that direction.  Wakes therefore evolve in both time and in the $(y,z)$ crossplane directions.  Assuming that the ship is traveling at a constant speed $U_0$, the $x$--coordinate location in the evolving wake can be reconstructed from the simulation time as $x=U_0t$.   

The computational domain is a rectangular parallelepiped, with dimensions, in meters, of $-500 < y < 500$ and $-500 < z < 0$.  Mesh resolution is approximately \unit[0.3]{m} with mesh density clustered near the ship and coarsening in the far field, with a total mesh size of 2,200,000 cells.   Time-stepping is set to $\Delta t = \unit[1]{s}$ and simulations are conducted over a period of \unit[1]{hour}.   The $x=\unit[0]{m}$ position is set as the initial data plane (IDP) and, so as to not violate the assumptions of the 2D+t approach, is located beyond the rapidly evolving near wake.  In a classical shear--flow, this would be at the point where asymptotic behavior begins.  For practical reasons, we assume this location to be about one--half of a ship length downstream from the stern.  
\subsection{Numerical scheme}
Solution of the governing equations is achieved using the segregated pressure--implicit split--operator (PISO) algorithm \citep{iss85}.  In this approach, the equations are sequentially solved, and the nonlinear advection term is linearized with velocity from the previous time step.  In our solver, the order of solution within a time step is as follows:  1) compute density given the temperature, salinity, and pressure fields from the previous time step; 2) solve the momentum equations for the velocity field, but with pressure from the previous time step; 3) iteratively solve the pressure--Poisson equation and velocity--field correction to enforce mass conservation; 4) solve the turbulence--model equations. 
The process is then repeated for each time step over the simulation.

The governing partial--differential equations are discretized using a cell--centered finite--volume method, which makes use of the Gauss--Green theorem which states that the surface integral of a scalar function is equal to the volume integral of the gradient of the scalar function.  Gradients at the cell faces are computed using weighted interpolation between the adjacent cell centers.  

To limit numerical dissipation, a second--order energy conserving scheme was used for the divergence operator in the advection term of equation \ref{eq:momRAS}.   Detailed studies were undertaken to properly select parameters that balance accuracy and stability, and to demonstrate over long time integration that the  prescribed kinetic energy, in the absence of turbulent eddy viscosity, could be maintained.  The scheme blends 2nd-order linear interpolation with a small amount of upwinding, and additionally applies a filter to eliminate high--frequency ringing.  All viscous Laplacian terms were discretized with a second--order linear scheme.  The time derivatives were discretized with a second--order backward finite difference.

The resultant algebraic system of equations are solved using iterative schemes.  The momentum and scalar--transport equations are solved using a pre--conditioned bi--conjugate gradient (PBiCG) scheme, with absolute and relative convergence criterion of $10^{-8}$ and $10^{-5}$, respectively. This means that the solution residuals must be less than $10^{-8}$ or be reduced 5 orders--of--magnitude over the iteration process.  The pressure-Poisson equation is solved with a generalized algebraic multi-grid (GAMG) scheme, with absolute and relative convergence criterion of $10^{-5}$ and $10^{-3}$, respectively.  For the pressure--velocity coupling in the PISO algorithm, two iterations are used; and since the mesh is perfectly orthogonal, zero non--orthogonal correctors are required.

To support our analysis, data is extracted from the simulations.   Locations of data extraction include one--dimensional lines on the ocean surface in the $y$--direction at every time step, where time $t$ can subsequently be converted to $x$ so that $(x,y)$ ocean--surface contour maps can be created. In addition, the entire $(y,z)$ cross plane is saved at various points in time that correspond to locations downstream of the ship. To visualize large--scale circulations (LTC), we present contours of the streamfunction $\psi$, which is computed from the $x$--component of the vorticity field $\omega_x$ by solving the Poisson equation, $\nabla^2\psi = -\omega_x$.  These contour maps serve as a good tool for understanding both the genesis and dynamics of LTC and LTC--induced surface currents, and the structure and longevity of the persistent wake.

\subsection{Initial and boundary conditions}
\begin{figure} 
\centering
\begin{overpic}[width=0.85\textwidth,tics=10]{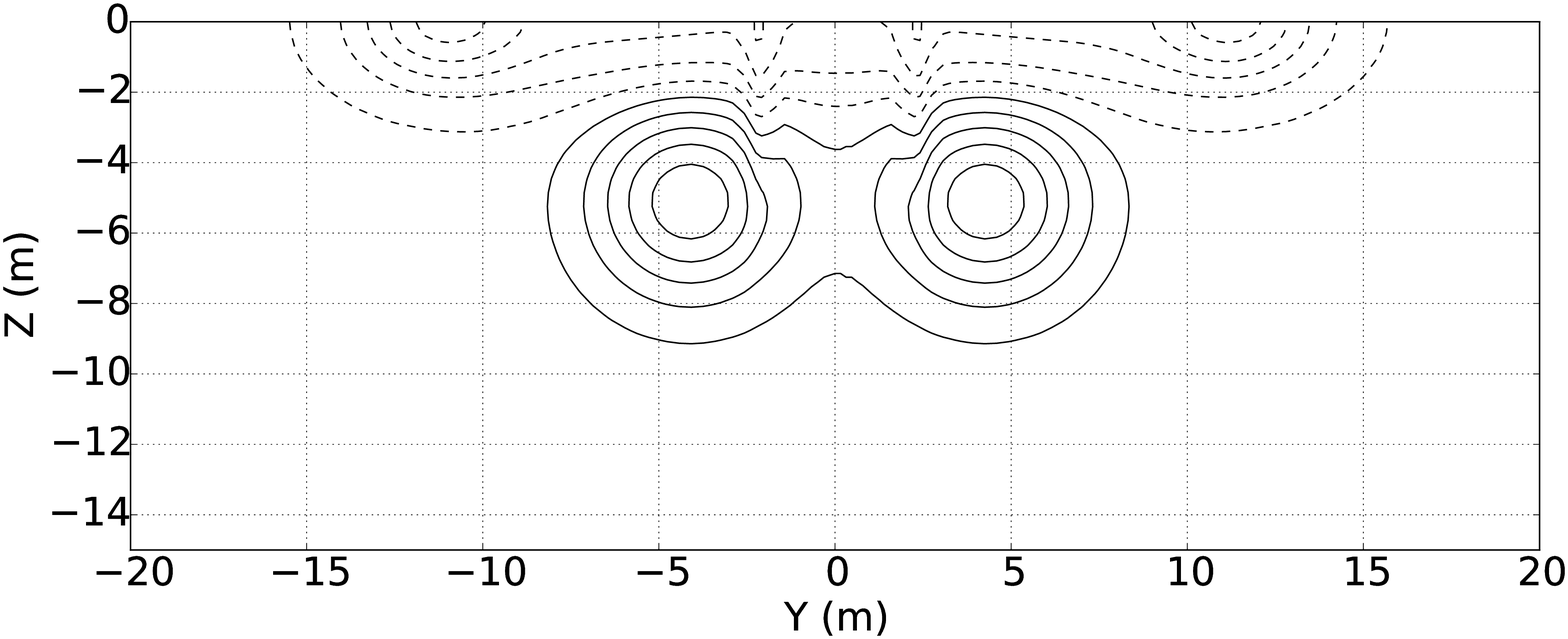}
 \put (12,37) {\large$\displaystyle (a)$}
\end{overpic} 
\begin{overpic}[width=0.85\textwidth,tics=10]{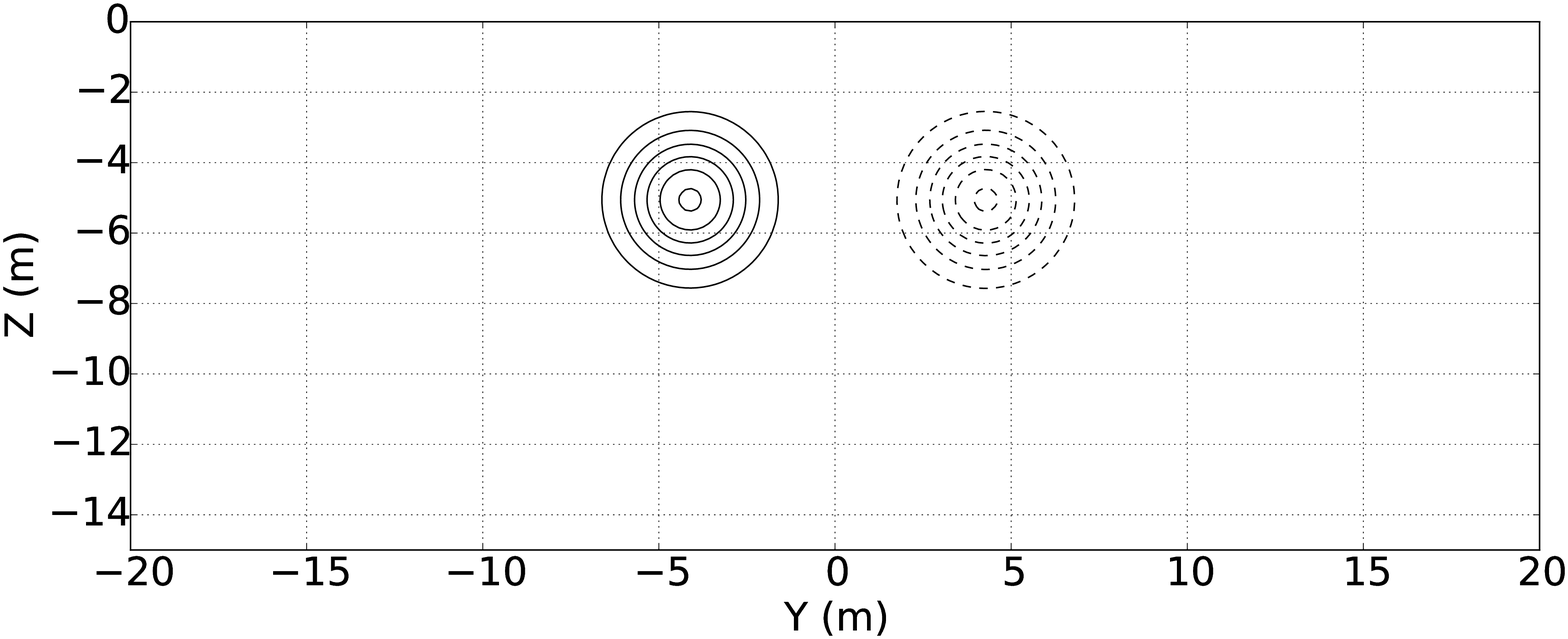}
 \put (12,37) {\large$\displaystyle (b)$}
 \end{overpic} 
  \caption{Initial data plane at $Fr = 0.35$. a) contour map of axial--component of velocity, where contour spacing is 1$\%U_0$, contour range is -4.95$\%U_0$ to 5.4$\%U_0$, and dashed lines represent negative contours.  b) contour map of streamfunction, where contour (streamline) spacing is 0.1 $m^2/s$.}
  \label{fig:IDP}
\end{figure}

Initial conditions at t=\unit[0]{s} must be specified at the indfitial data plane (IDP) for all simulation variables, $U, k, \epsilon, T, S$, and $p$. The IDP can be prescribed from theory, geometry--resolving CFD simulations, or even experimental data.  For the work here, an empirical--analytical representation of the wake one-half ship length downstream, figure \ref{fig:IDP}, is established based on a formulation similar to \cite{min88}.  This model includes resistance contributions from the ship's hull, wave breaking, and rudders; in addition to the thrust current and swirl from the propellers as detailed in  Appendix A.

The axial velocity produced by the ship is presented in figure \ref{fig:IDP}a as a contour plot of its magnitude on the $(y,z)$ plane.  The axial velocity is positive in the direction of the thrust current, which is opposite to the direction of ship velocity.  The strongest current is produced obviously in the area of the propellers.  We call it a thrust current, whereas a negative current in the direction of the ship velocity we call the drag current since it is created by the drag of the ship hull.  The velocity is normalized by the ship speed $U_0$.  Of course, for all other factors being equal, the ratio between the velocity magnitudes of thrust and drag currents depends on the ship speed and, correspondingly, on Froude number.  At higher speeds, the contribution of wave--making resistance  increases and the relative contribution of frictional resistance decreases.

The transverse (rotational) flow produced by the ship propellers is presented in figure \ref{fig:IDP}b as contour lines of the streamfunction on the same $(y,z)$ plane.  The contour map of streamfunction in figure \ref{fig:IDP}b should not be confused with the contour lines of the magnitude of axial velocity used in the previous figure  \ref{fig:IDP}a.  The map of the streamfunction shows the direction of the transverse velocity.  Proximity of the streamlines to each other (density) characterizes the magnitude of the rotational velocity.  Knowing that every contour in figure \ref{fig:IDP}b corresponds to a change in streamfunction of 0.1 $m^2/s$, one can estimate the velocity.  This representation of the transverse flow as streamlines is employed throughout this paper to characterize the evolution of circulations in the ship wake in the manner used by \cite{bas11}.  Since the propellers in this case are rotating outward, the circulatory motion  represented in figure \ref{fig:IDP}b is also outward rotating.  The maximum computed value of transverse velocity of the flow presented in figure \ref{fig:IDP}b is below the surface and has a value of 0.4 m/s.


To perform 2D+t simulations, periodic boundary conditions are used in the axial $x$--direction for all variables.  The ocean surface is treated as a slip boundary (zero gradient for all variables, except for the normal component of velocity which is set to zero) to allow the formation of transverse surface currents; the bottom is also treated as a slip boundary, but is set sufficiently below the area of interest so as to not influence mass conservation and the pressure field; the far--field side boundaries are treated as pressure--outflow boundaries, but are also set sufficiently wide to prevent internal--gravity wave reflections when running stratified simulations.

\subsection{Fluid properties and flow conditions}
The fluid properties are described by the density and the kinematic viscosity.   In general, the density in our solver is computed using the TEOS-10 seawater equation--of--state.  To isolate the effects of the LTC, and since surface ships often operate in the ocean mixed layer, it is assumed here that the fluid is iso--thermal and iso--haline with $T=\unit[20]{C}$ and $S=\unit[35]{psu}$.  Furthermore, to eliminate the generation of weak internal gravity waves we have disabled the influence of pressure on the equation of state, thus creating an isopycnic fluid with a density of $\rho=\unit[1025]{kg/m^3}$.  The kinematic viscosity is set to $\nu=\unit[1.0\times10^{-6}]{m^2/s}$. 

Surface wave parameters are set to realistic ocean conditions with wave amplitudes of 0 and \unit[0.25]{m} and wavelength of \unit[10]{m}.  Relative wave--ship heading angles of $0^\circ$ and $180^\circ$ are used to simulate head-- and following--seas conditions.  Full--scale ship parameters are used with $Fr=0.35$, $Re=1.85\times10^{9}$, and a ship speed of $U_0=\unit[13]{m/s}$.  

Simulations are conducted for an IDP that would be generated by a full--scale naval surface combatant, as described by the David Taylor Model Basin (DTMB) Model 5415.  The hull geometry includes both a sonar dome and transom stern, and propulsion is provided through twin open--water propellers driven by shafts supported by struts.  For the full--scale ship, the key geometric parameters which serve as inputs to our theoretical model, described in Appendix A, are as follows; length $L = \unit[142]{m}$, beam $B=\unit[18.9]{m}$, draft $D=\unit[6.16]{m}$, displacement $\nabla = \unit[8425.4]{m^3}$, wetted surface area $S=\unit[2949.5]{m^2}$, propeller diameter $D_p = \unit[5.29]{m}$, propeller depth $z_p = \unit[5.11]{m}$, propeller horizontal offset $y_{p}=\unit[4.2]{m}$, rudder depth $z_r = \unit[1.76]{m}$, rudder horizontal offset $y_r=\unit[3.16]{m}$, rudder thickness $t_r =\unit[0.95]{m}$, and rudder planform area $S_r=\unit[12.1]{m^2}$. 

\section{Simulations of wake features under different surface wave conditions }
\subsection{Wake evolution without ambient surface waves (calm seas)}

First, we consider the case of wake evolution without ambient surface waves.  This case is important since it allows us to evaluate the characteristic lifetime of the wake without interaction of the ship--generated current with the surface waves, and later compare it to the persistent wake caused by such interaction.

\begin{figure}
	\centering
    \begin{overpic}[width=0.8\textwidth,tics=10]{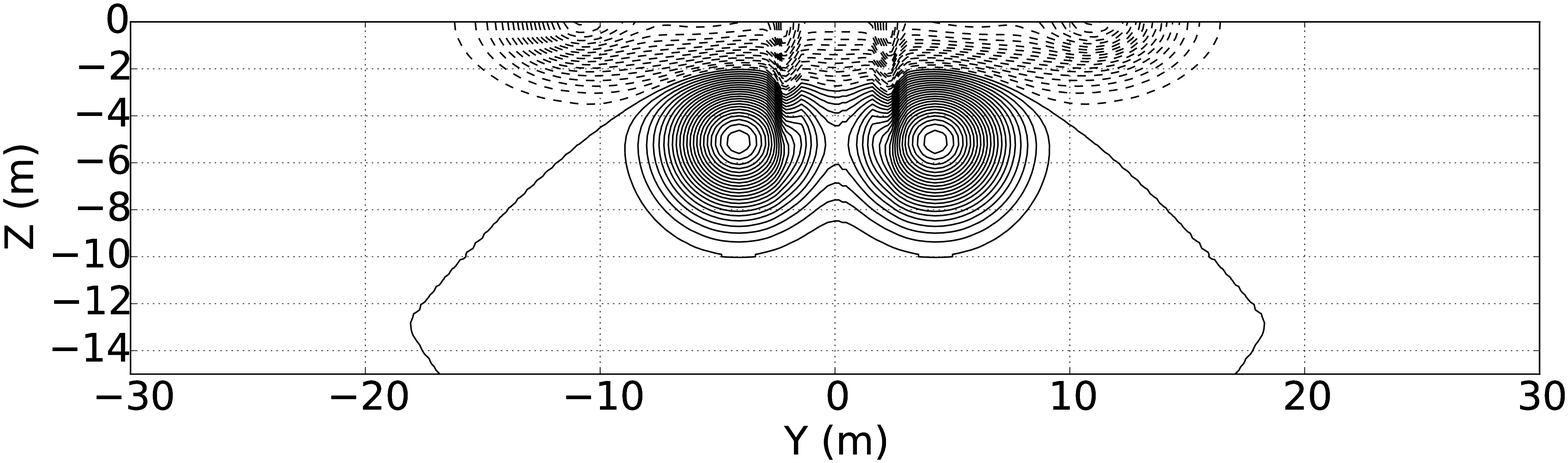}
	 \put (12,26) {\large$\displaystyle (a)$}
	\end{overpic} 
    \begin{overpic}[width=0.8\textwidth,tics=10]{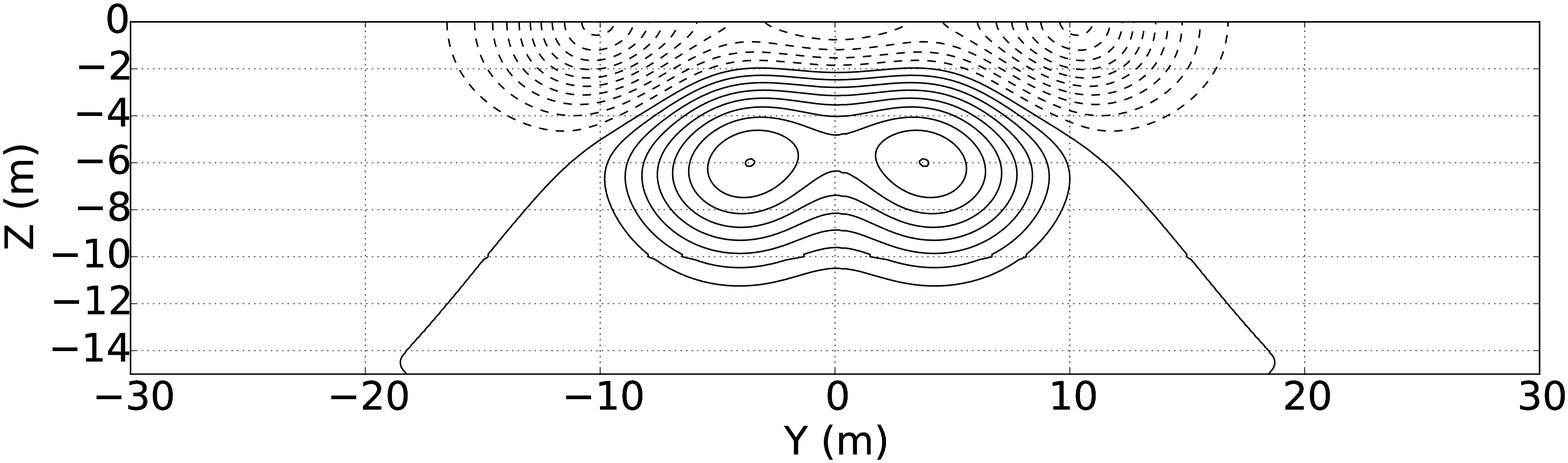}
	 \put (12,26) {\large$\displaystyle (b)$}
	\end{overpic} 
     \begin{overpic}[width=0.8\textwidth,tics=10]{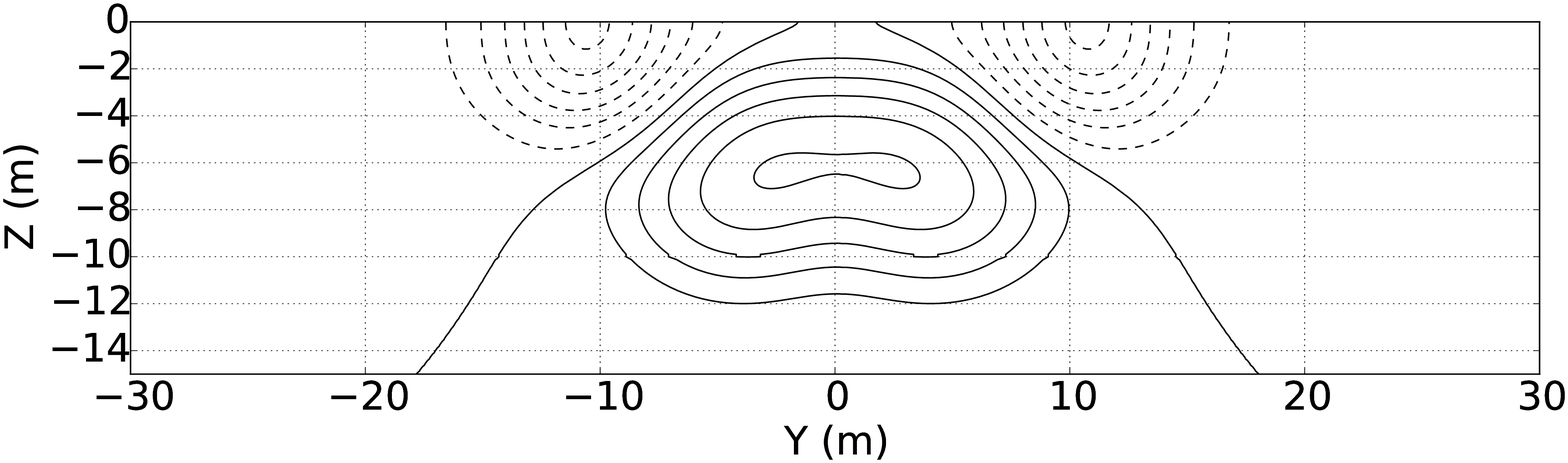}
	 \put (12,26) {\large$\displaystyle (c)$}
	\end{overpic} 
     \begin{overpic}[width=0.8\textwidth,tics=10]{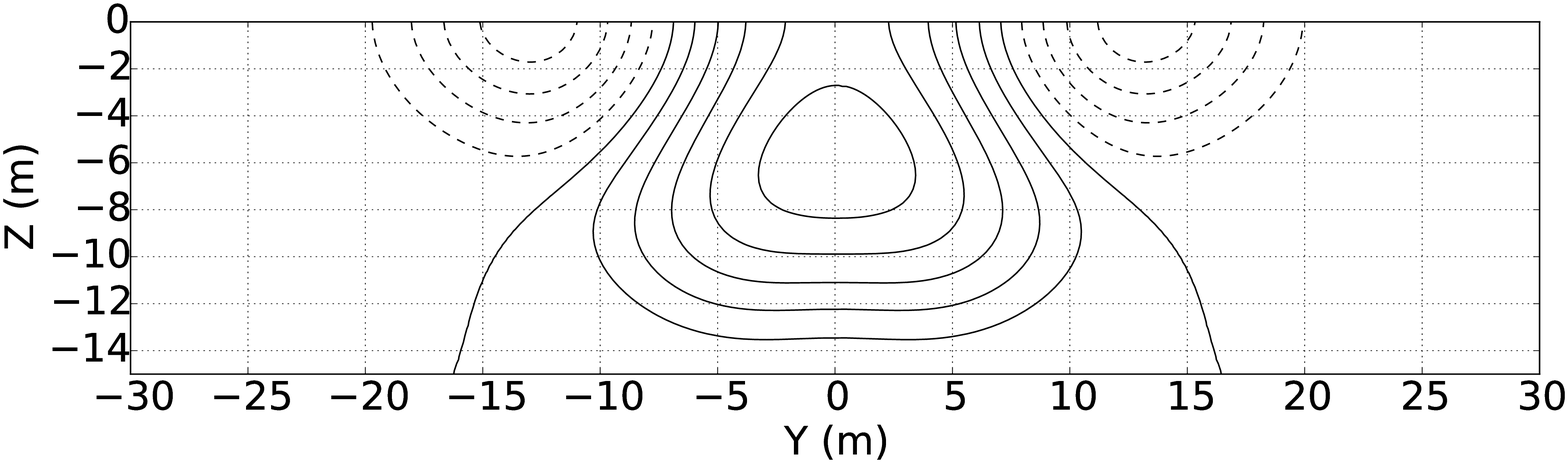}
	 \put (12,26) {\large$\displaystyle (d)$}
	\end{overpic} 
    \caption{Contours of axial velocity at (a) $x = \unit[100]{m}$, (b) $x =$ \unit[450]{m}, (c) $x =$ \unit[850]{m}, and (d) $x =$ \unit[1250]{m} at $Fr=0.35$, without ambient surface waves. Contour spacing = \unit[0.02]{m/s}.  Dashed lines are negative contours.}  
    \label{fig:axialContoursCalm}
\end{figure}

The evolution of the IDP (figure \ref{fig:IDP}) is shown in figure \ref{fig:axialContoursCalm} through snapshots of the velocity component along the $x$--axis at four downstream distances (100m, 450m, 850m, and 1250m).  The four snapshots illustrate the evolution of the axial current: the two separate thrust currents produced by the propellers  (shown in figure \ref{fig:axialContoursCalm}a) merge relatively soon into one larger jet, in agreement with \cite{hym98}.  By a distance of 850m behind the ship, the drag current has the appearance of two smaller jet--like currents above and on each side of the thrust current  (figure \ref{fig:axialContoursCalm}c), but with axial velocity in the opposite direction.  An awareness and the characterization of the axial flow component is crucial to understanding the generation of LTC, which result from the interaction of this axial flow with the surface wave field, as described later. The evolution of the transverse current and related streamlines is not shown for the case without ambient surface waves, since the diverging circulatory motions shown in figure \ref{fig:IDP}b simply decay as the distance behind the ship increases.  

\begin{figure}
	\centering
    \begin{overpic}[width=0.49\textwidth,tics=10]{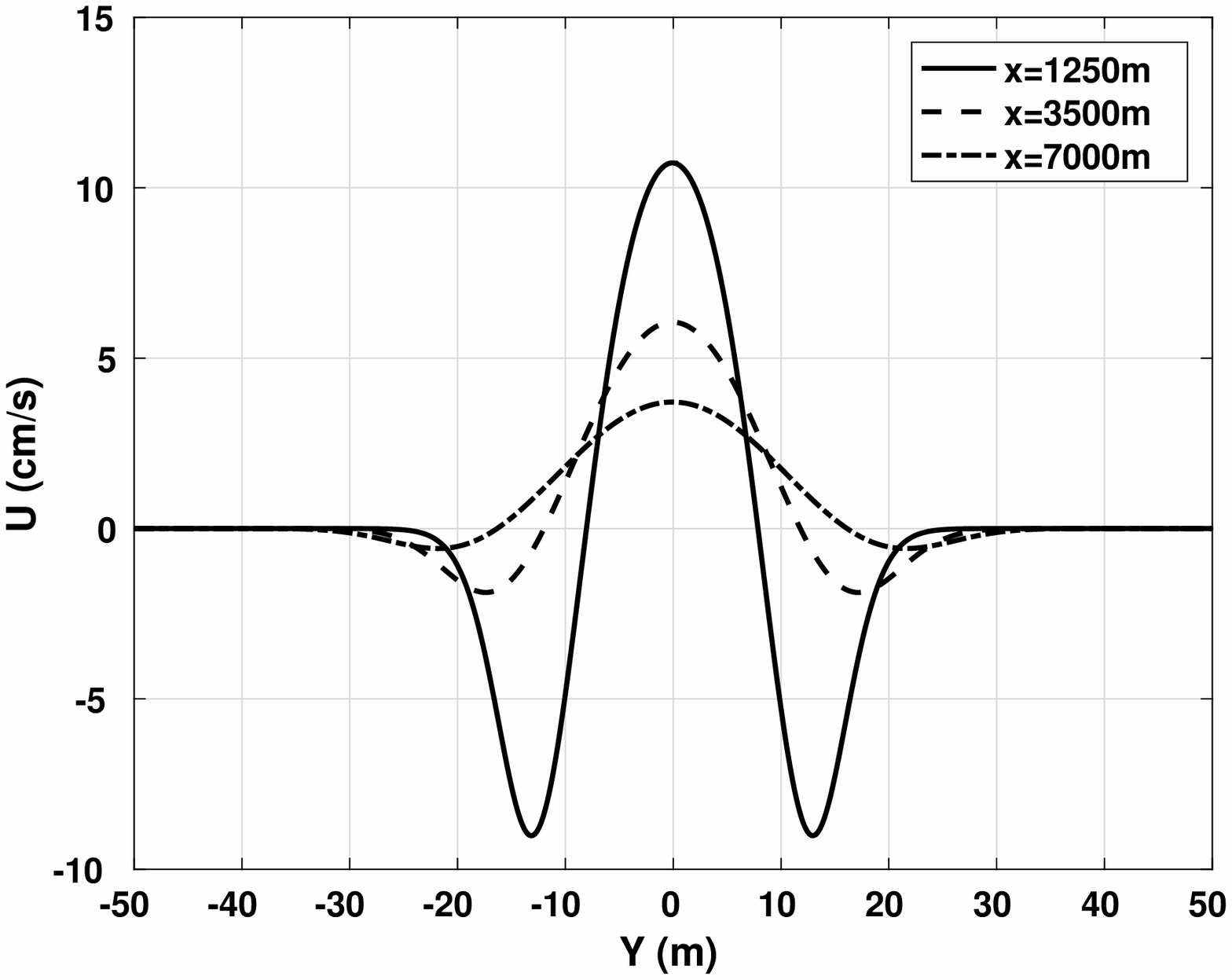}
	\put (11,70) {\large$\displaystyle (a)$}
	\end{overpic} 
	\begin{overpic}[width=0.49\textwidth,tics=10]{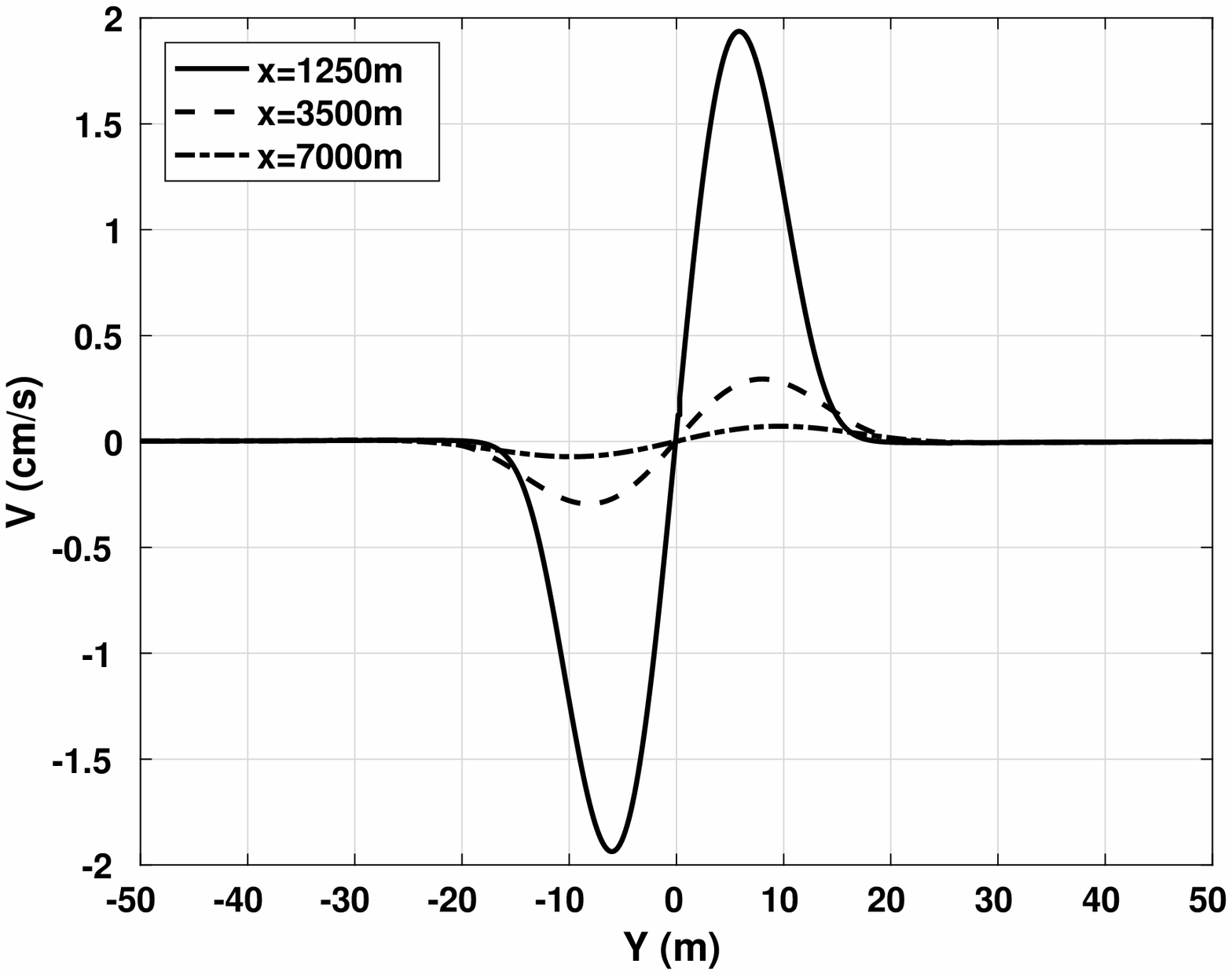}
	\put (90,70) {\large$\displaystyle (b)$}
	\end{overpic} 
	\caption{Evolution of surface currents at $Fr=0.35$, without ambient surface waves.  a) axial--component of velocity; b) transverse--component of velocity.}
    \label{fig:currentProfilesCalm}
\end{figure}

\begin{figure}
	\centering
    \begin{overpic}[width=8.5cm]{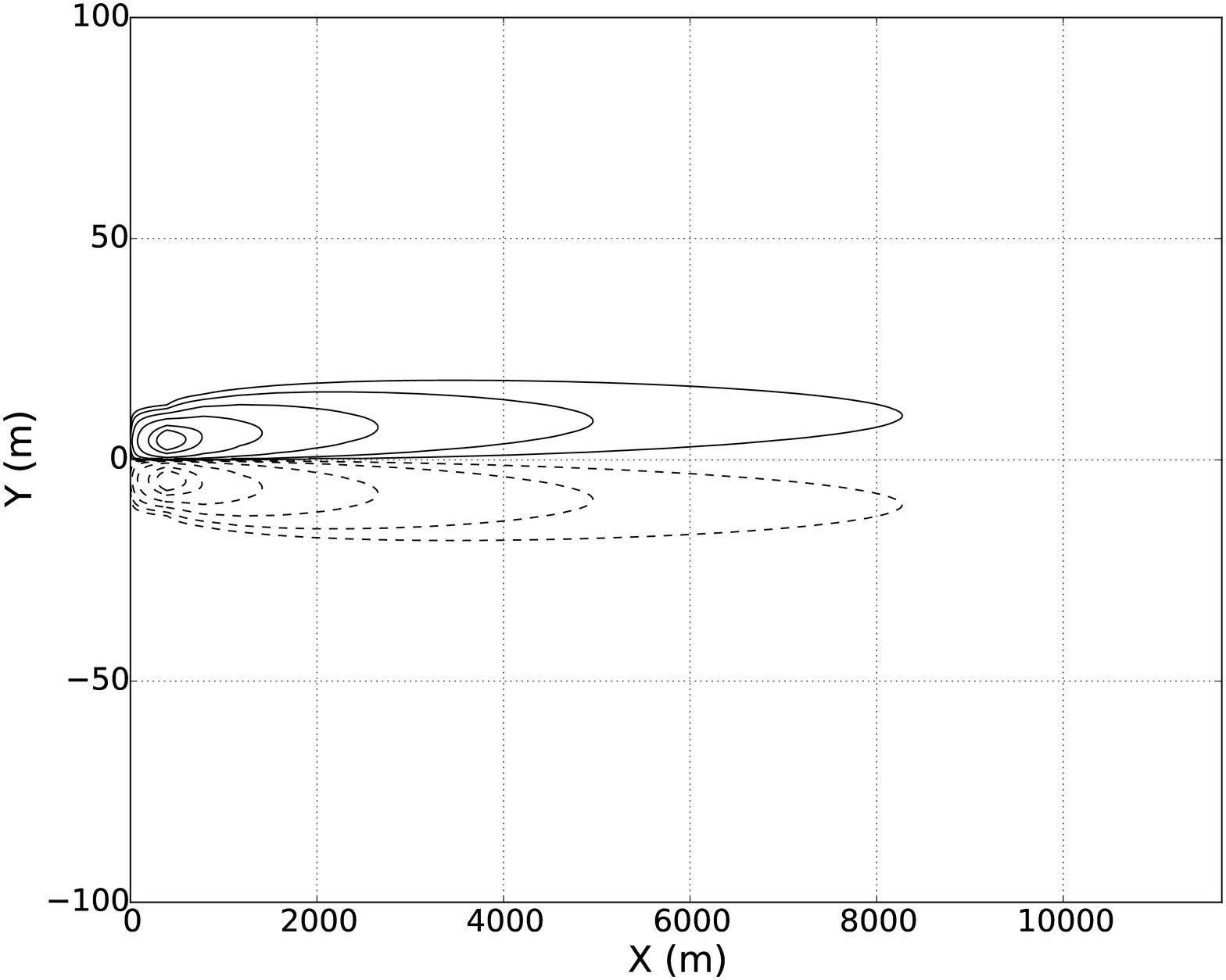}
    \put (45,58) {\vector(-0.5,-1.4){2.8}}
      \put (45,58) {\vector(0.5,-1.4){8}}
     \put (40,62) {\colorbox{white}{\parbox{0.1\linewidth}{Converging Zones}}}  
     \put (30,30) {\vector(0,1){13}}
      \put (23,23) {\colorbox{white}{\parbox{0.1\linewidth}{Diverging Zone}}} 
     \end{overpic} 
    \caption{Contour map of transverse--component of velocity on ocean surface at $Fr=0.35$ without ambient surface waves.  Contours = +/- 0.05, 0.15, 0.5, 1.5, 3.5, 5 cm/s.  Dashed lines are negative contours.}
    \label{fig:currentContoursCalm}
\end{figure}

In the present study of the persistent wake, we are primarily interested in the character of the near--surface currents, because they are responsible for the redistribution of SAS films at the surface  \citep{pel92,tal97}.   Figure \ref{fig:currentProfilesCalm}a  shows the distribution of the axial component and \ref{fig:currentProfilesCalm}b shows the transverse component of the surface velocity as a function of the transverse $y$--coordinate at three distances (1250m, 3500m, and 7000m) behind the ship.   Figure \ref{fig:currentContoursCalm} shows the distribution of the transverse velocity of the surface current on the $(x,y)$ plane in the frame of reference related to the ship, in which the flow is stationary.

As can be seen from figure \ref{fig:currentProfilesCalm}a, the axial surface velocity at each of the three distances behind the ship has regions with positive or negative values, which correspond to the thrust or drag currents, respectively.  The transverse flow, which is shown in figure \ref{fig:currentProfilesCalm}b, has a divergence zone in the middle of the wake (centerline), since the velocity is directed away from the centerline and the two weak convergence zones at the edges.  We call these convergence zones weak because the velocity on one side of such a zone is zero.  The flow shown in figure \ref{fig:currentProfilesCalm}b will redistribute SAS away from the centerline toward the edges.  Both the axial and transverse components of the velocity decrease with distance behind the ship.  The transverse component of the surface velocity decays faster than does the axial.  Figure \ref{fig:currentContoursCalm} illustrates that the transverse velocity practically disappears at a distance on the order of 5000m behind the ship, which is about 35 ship lengths.  

\subsection{Wake evolution in presence of ambient surface waves}

Two effects generated by the ship, near--surface turbulence and surface flow, are mainly responsible for the ocean--surface manifestations of the ship wake.  They each affect the short surface waves that define SAR and optical images of the surface.  Turbulence increases dissipation of the short waves directly.  Surface currents change the intensity of the short surface waves by changing their propagation velocity and by changing SAS concentration at the surface, which in turn affect short--wave dissipation.  The correct description of the relative contributions of these two mechanisms that alter the short surface waves is still a work in progress.  However, some qualitative conclusions about such contributions can be made.  The direct kinematic effect of the surface currents on the short surface waves can be significant in the near-- and  far--wake regions, where the ship--induced surface current is still relatively strong.  If the effect of the surface currents is dominant, the change in the short surface--wave field is asymmetric: the change is more pronounced on one side of the wake \citep{fuj16}.  For the weak currents found in the persistent wake, redistribution of SAS at the surface may be more important than the change of kinematics of the surface waves, especially if the initial concentration of SAS is significant.  This redistribution is mainly caused by the transverse component of the surface current when the SAS films are collected in convergence zones.  When the effect of the SAS films is dominant, then the change in the amplitude of the short waves is expected to be symmetric, as observed in figures \ref{fig:HeadSeaSAR} and \ref{fig:FollowingSeaSAR}.

The genesis of the persistent wake and its structure are described in the next sections.  The ship--induced persistent wake develops in the presence of ambient surface waves that interact with the ship--generated current described above.  This interaction can produce large--scale LTC that exist for an extended time and, correspondingly, a long distance behind the ship.  The vorticity in LTC and the structure of the persistent wake depend on the relative direction between the surface waves and the ship course \citep{bas11}.  Therefore, we consider two cases: first, the ship is headed into the seas, that is the ship travels in a direction generally opposite to the direction of propagation of the surface waves (head seas) and, second, the ship's course is in the same general direction as the direction of the surface waves (following seas).  In the head seas case, the angle between the ship's velocity vector and the wave vector of the surface waves is greater than 90 degrees, while for the following seas case this angle is less than then 90 degrees.  When this angle is close to 90 degrees, the vortex force imposed on the initial ship-generated current is close to zero.  Thus, for the 90 degree case, the LTC are not generated and there is no persistent wake.  First the case of head seas is considered, since in this case, the transition from far wake to persistent wake is easier to observe.
\subsection{Wake evolution for ship in head seas}
In this section we consider simulations of the persistent wake for the case of head seas.  The amplitude and wavelength of the surface waves have been selected based on the wind speed and fetch described in the paper on full--scale at--sea experiments by \cite{mil93}.  For simplicity, we consider a sinusoidal wave with amplitude of 0.25 m and wavelength of 10m.   It should be noted that an exact characterization of the surface wave spectrum is not crucial to the success of simulations.  It is important that the vortex force is present and the direction of propagation of surface waves relative to the ship course, which defines the structure of the persistent wake, is known.  
   
\begin{figure}
	\centering
\begin{overpic}[width=0.75\textwidth,tics=10]{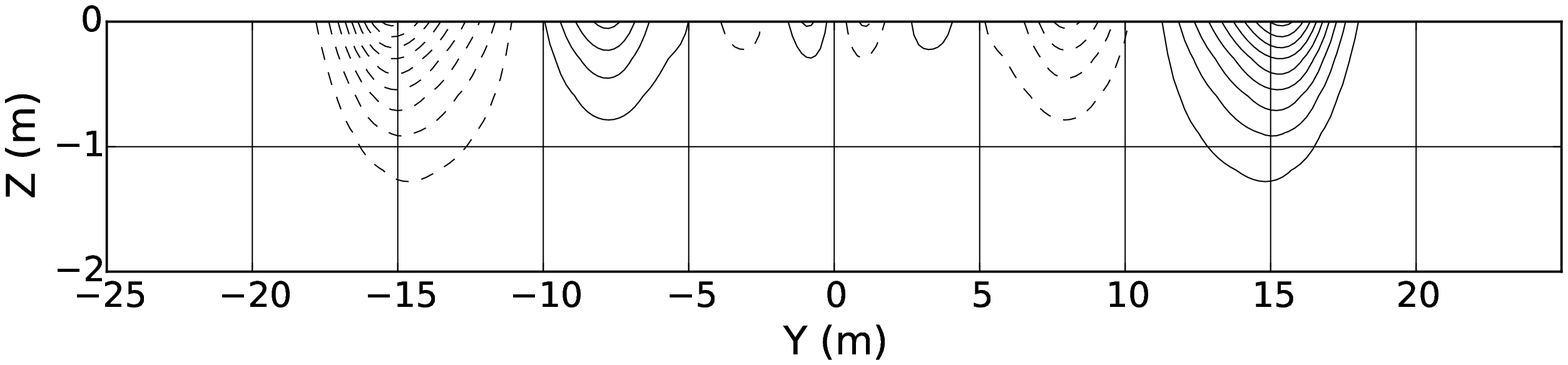}
\put (10,18) {\large$\displaystyle (a)$}
\end{overpic} 
\begin{overpic}[width=0.75\textwidth,tics=10]{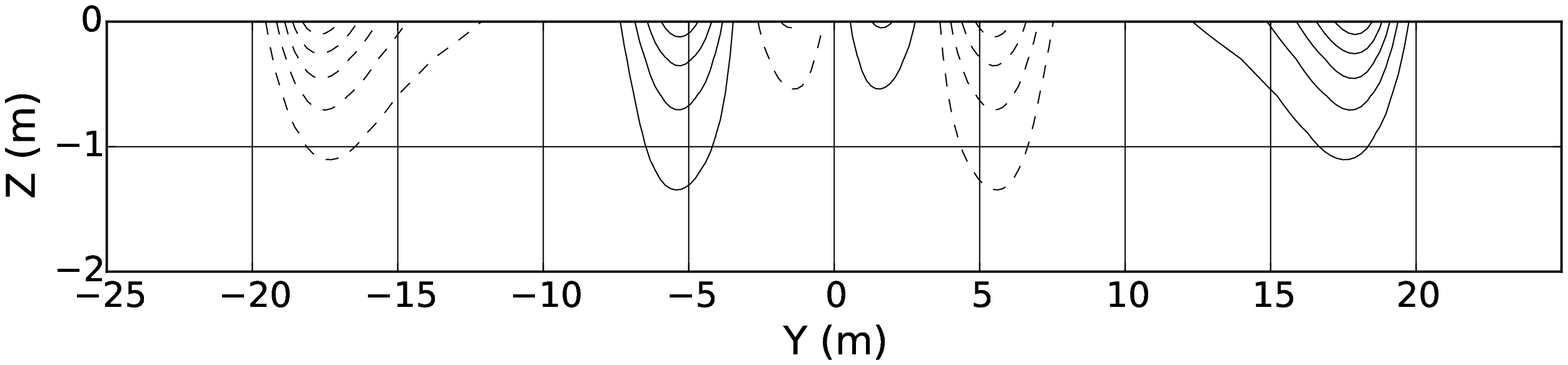}
\put (10,18) {\large$\displaystyle (b)$}
\end{overpic} 
\begin{overpic}[width=0.75\textwidth,tics=10]{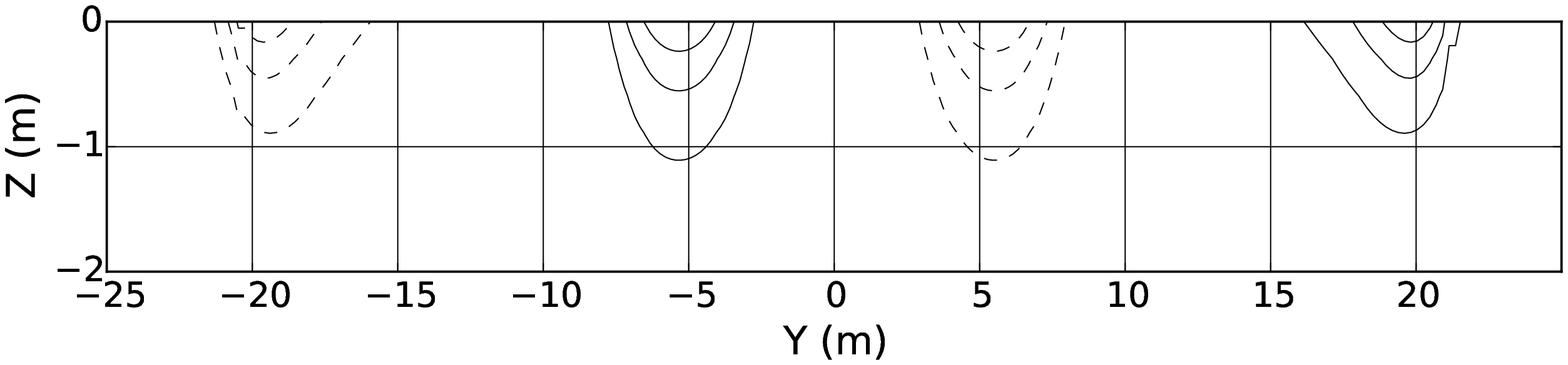}
\put (10,18) {\large$\displaystyle (c)$}
\end{overpic}
\caption{Contour maps of the horizontal component of vortex force $F_y$ at (a) $x$ = 450m, (b) $x =$ 850m, and (c) $x =$ 1250m in head seas at Fr = 0.35.  Surface waves with $\lambda = 10m$ and $a_s = 0.25m$.  Contour spacing = 0.001 N/kg.  Peak Vortex Force at x=450m is 0.0094 N/kg.  Dashed lines are negative contours.}
\label{fig:VF_Y}
\end{figure}
\begin{figure}
\centering
\begin{overpic}[width=0.75\textwidth,tics=10]{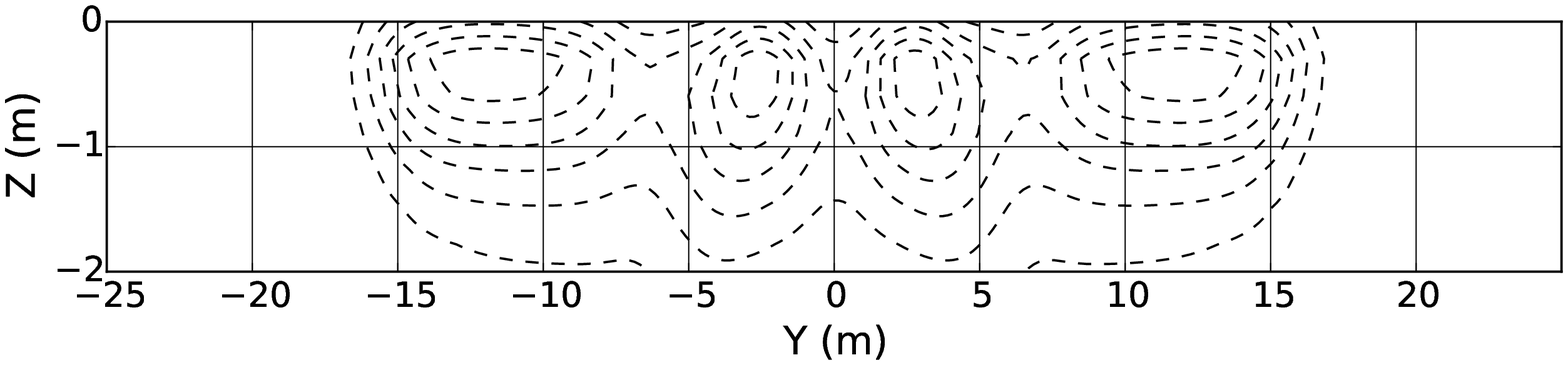}
\put (10,18) {\large$\displaystyle (a)$}
\end{overpic}
\begin{overpic}[width=0.75\textwidth,tics=10]{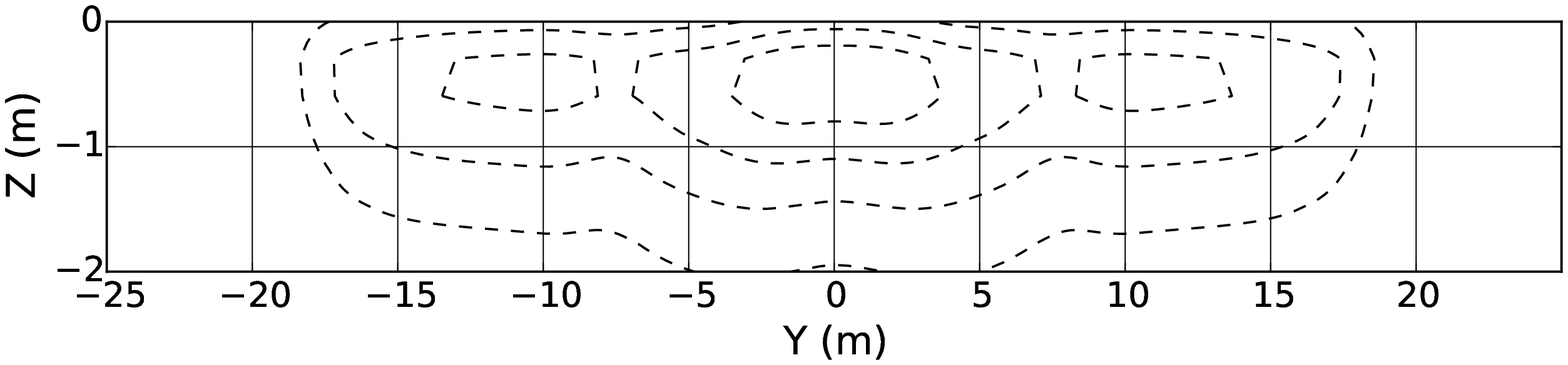}
\put (10,18) {\large$\displaystyle (b)$}
\end{overpic}
\begin{overpic}[width=0.75\textwidth,tics=10]{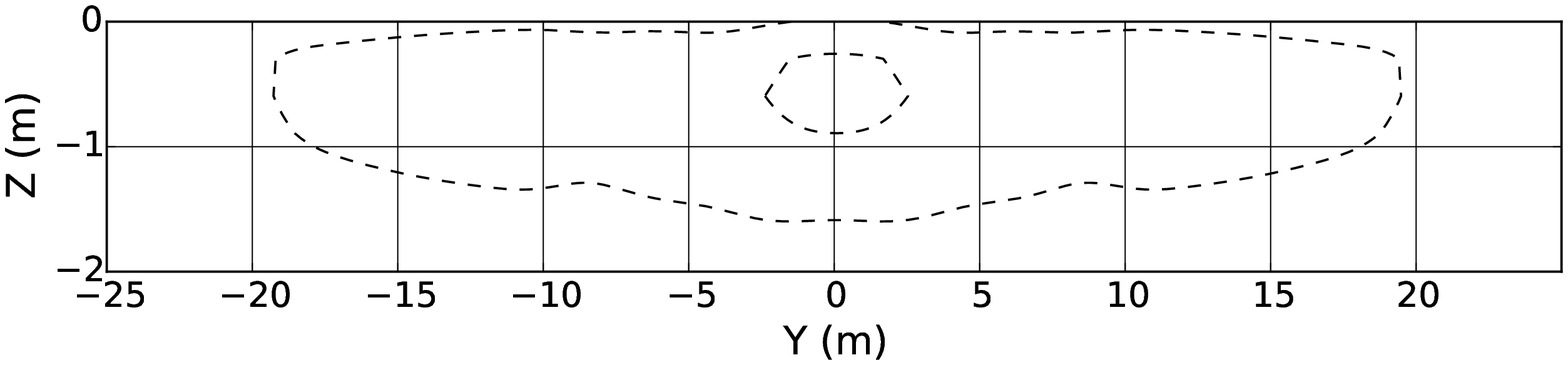}
\put (10,18) {\large$\displaystyle (c)$}
\end{overpic}
\caption{Contour maps of the vertical component of vortex force $F_z$ at (a) $x$ = 450m, (b) $x =$ 850m, and (c) $x =$ 1250m in head seas at Fr = 0.35.  Surface waves with $\lambda = 10m$ and $a_s = 0.25m$.  Contour spacing = 0.001 N/kg.  Dashed lines are negative contours.}
\label{fig:VF_Z}
\end{figure}

\begin{figure}
\centering
\begin{overpic}[width=0.65\textwidth,tics=10]{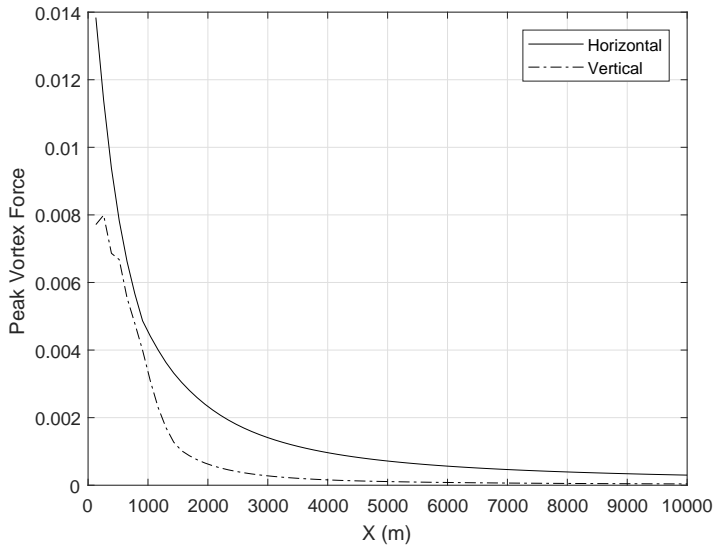}
\end{overpic}
\caption{Decrease of maximum (peak) values of the components of vortex force (N/kg) in head seas as a function of distance behind the ship at Fr = 0.35.  Surface waves with $\lambda = 10m$ and $a_s = 0.25m$.}
\label{fig:Head_VF}
\end{figure}

First, the distribution of the vortex force and its change with the distance behind the ship are considered. Figures \ref{fig:VF_Y} and \ref{fig:VF_Z} show the evolution of horizontal and vertical components of the vortex force in head seas.  The contour lines are used to describe distribution of the vortex force components which are shown for three different distances behind the ship.  At relatively small distance (Fig.7 a) the structure of the horizontal component of  the vortex force is complex with several regions where the force changes sign. This structure reflects complexity of the initial current which is in turn influenced by the ship geometry.  As the structure of the ship induced drag current bifurcates, the distribution of the horizontal component of the vortex force simplifies into four symmetric regions with alternating directions and increasing distance between them.  The vertical component of the vortex force maintains a primarily negative direction as the vertical velocity gradient is negative nearest the surface.  Distribution of the vertical component of the vortex force changes in a similar manner to the horizontal component, but decays faster than the horizontal component.  As the ship induced current evolves (see figure \ref{fig:axialContoursCalm}) the horizontal velocity gradient near the surface becomes greater than the vertical gradient, leading to a stronger and more persistent horizontal component of the vortex force than vertical (see figure \ref{fig:Head_VF}).

\begin{figure}
\centering
\begin{overpic}[width=0.65\textwidth,tics=10]{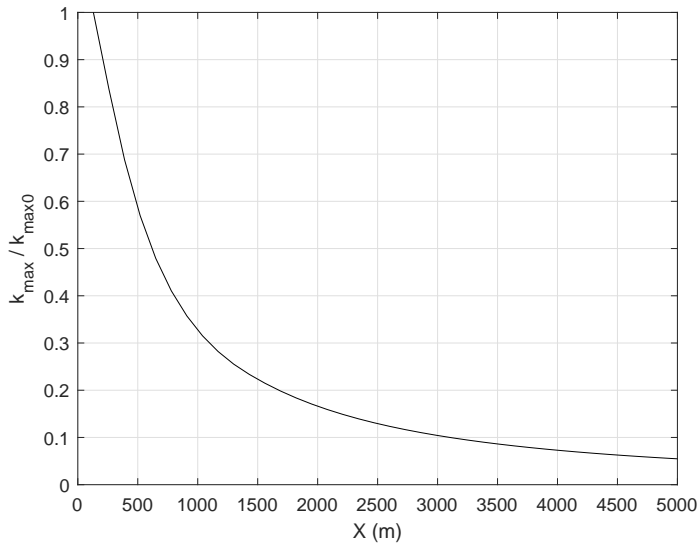}
\end{overpic}
\caption{Decay of the maximum (peak) value of the turbulent kinetic energy as a function at Fr=0.35.  Values are non-dimensionalized by the maximum initial value of the TKE (k$_{max0}$).}
\label{fig:Peak_TKE}
\end{figure}

The horizontal component of the vortex force creates converging and diverging flows along the surface with spacing between them on the order of tens of meters.  Converging  flows form downwelling regions and diverging flows form upwelling regions due to continuity, which lead to the formation of large scale LTC.  These circulations, once formed, can persist for tens of kilometers, as the ship induced turbulence decays as shown in figure \ref{fig:Peak_TKE}.  Maximum TKE is observed to have reduced by an order of magnitude around the distance of \unit[3000]{m}.

\begin{figure}
	\centering
    \begin{overpic}[width=0.75\textwidth,tics=10]{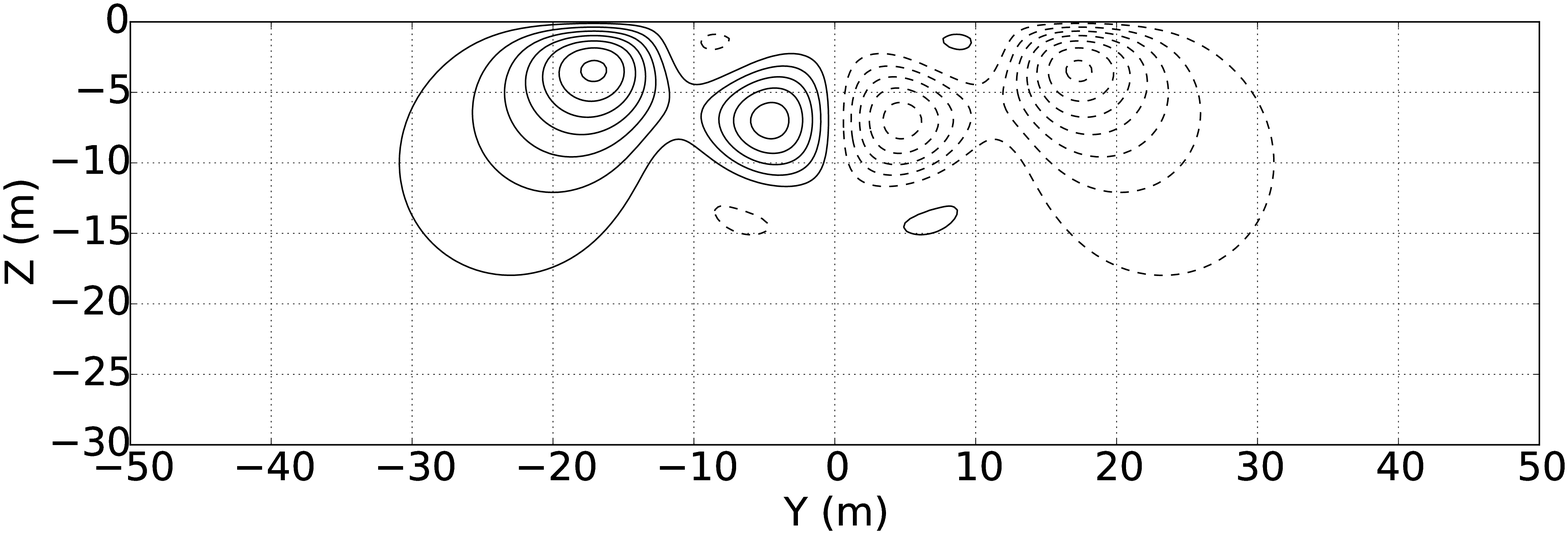}
	 \put (11,30) {\large$\displaystyle (a)$}
     \put (53.5,15) {\vector(-0.35,1){3.7}}
     \put (53.5,15) {\vector(0.35,1){3.7}}
     \put (50.5,11.5) {\colorbox{white}{\parbox{0.02\linewidth}{(1)}}}
	\end{overpic} 
        \begin{overpic}[width=0.75\textwidth,tics=10]{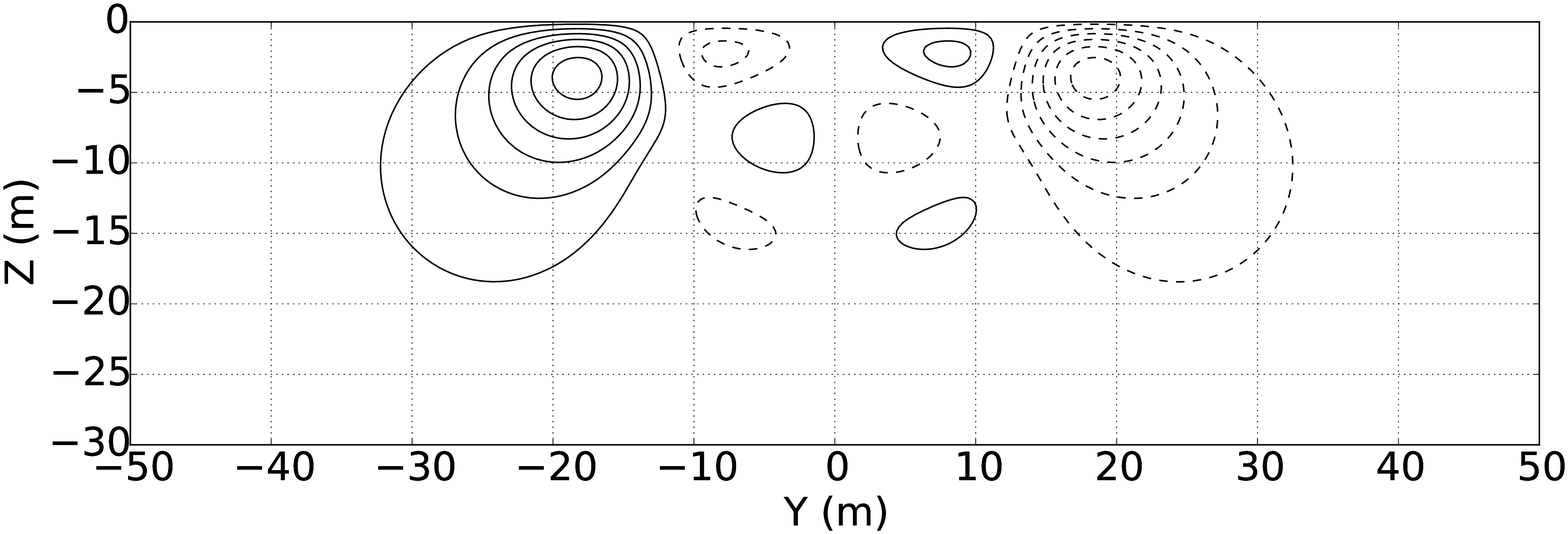}
	 \put (11,30) {\large$\displaystyle (b)$}
     \put (53.5,15) {\vector(-0.35,1){3.7}}
     \put (53.5,15) {\vector(0.35,1){3.7}}
     \put (50.5,11.5) {\colorbox{white}{\parbox{0.02\linewidth}{(1)}}}
	\end{overpic} 
     \begin{overpic}[width=0.75\textwidth,tics=10]{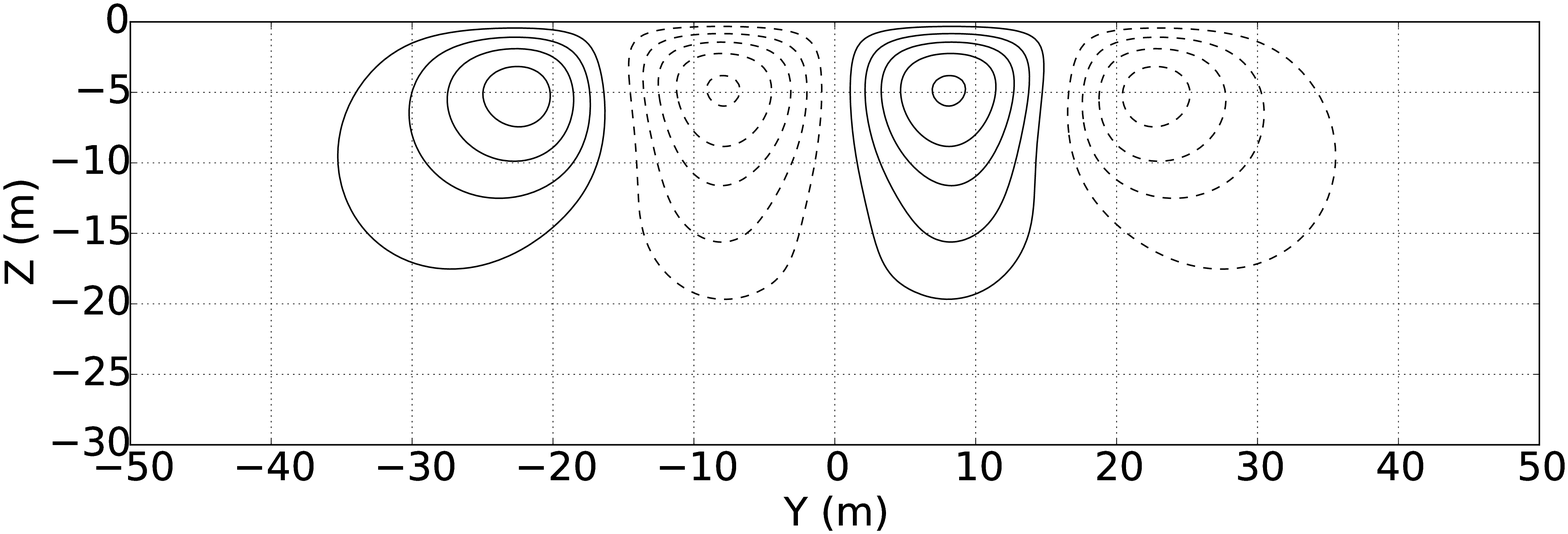}
	 \put (11,30) {\large$\displaystyle (c)$}
	\end{overpic} 
    \caption{Contour maps of streamfunction at (a) $x = \unit[1250]{m}$, (b) $x =$ \unit[1650]{m}, and (c) $x =$ \unit[3500]{m} in head seas at $Fr=0.35$.  Surface waves with $\lambda=\unit[10]{m}$ and $a_s=\unit[0.25]{m}$. Contour spacing = $\unit[0.005]{m^2/s}$.  Dashed lines are negative contours.  (1) designates twin propeller swirl, which is non-traceable at 3500m.  All other circulations are Langmuir-type circulations.}
    \label{fig:streamfunctionContoursHead}
\end{figure}

The development of the transverse flow in the ship wake for the case of head seas is shown in figure \ref{fig:streamfunctionContoursHead} for three different distances behind the ship.  These distances are chosen purely for illustration purposes and are within the range of values that are shown in figure \ref{fig:currentProfilesCalm}.  The structure of the current is very different from the case with no surface waves.  The initial distribution of the transverse velocity immediately behind the ship is the same as shown in figure \ref{fig:IDP}: two circulatory flows created by the ship propellers rotating outward.  These flows caused by propellers are still observed  in figure \ref{fig:streamfunctionContoursHead} several hundred meters behind the ship  (with their intensity reduced).  Simultaneously, four new LTC develop as a result of the vortex forcing as is seen in figure \ref{fig:streamfunctionContoursHead}a.  As the distance behind the ship increases, the circulatory motions caused by the propellers dissipate (see figure \ref{fig:streamfunctionContoursHead}b).  Finally they disappear completely by \unit[3500]{m} (figure \ref{fig:streamfunctionContoursHead}c) and only four circulations exist in the area far behind the ship creating the persistent wake.  Note, that the complex structure of the flow in the center of figure \ref{fig:streamfunctionContoursHead}a and figure \ref{fig:streamfunctionContoursHead}b is a result of superposition of flow caused by propellers and the inner set of developing LTC.   The developed structure of the LTC in the persistent wake for the head seas case, as presented in figure \ref{fig:streamfunctionContoursHead}c, is as follows: the inner circulations are inward rotating, relative to the surface centerline, while outer circulations are outward rotating. 

\begin{figure}
\centering
\begin{overpic}[width=0.75\textwidth,tics=10]{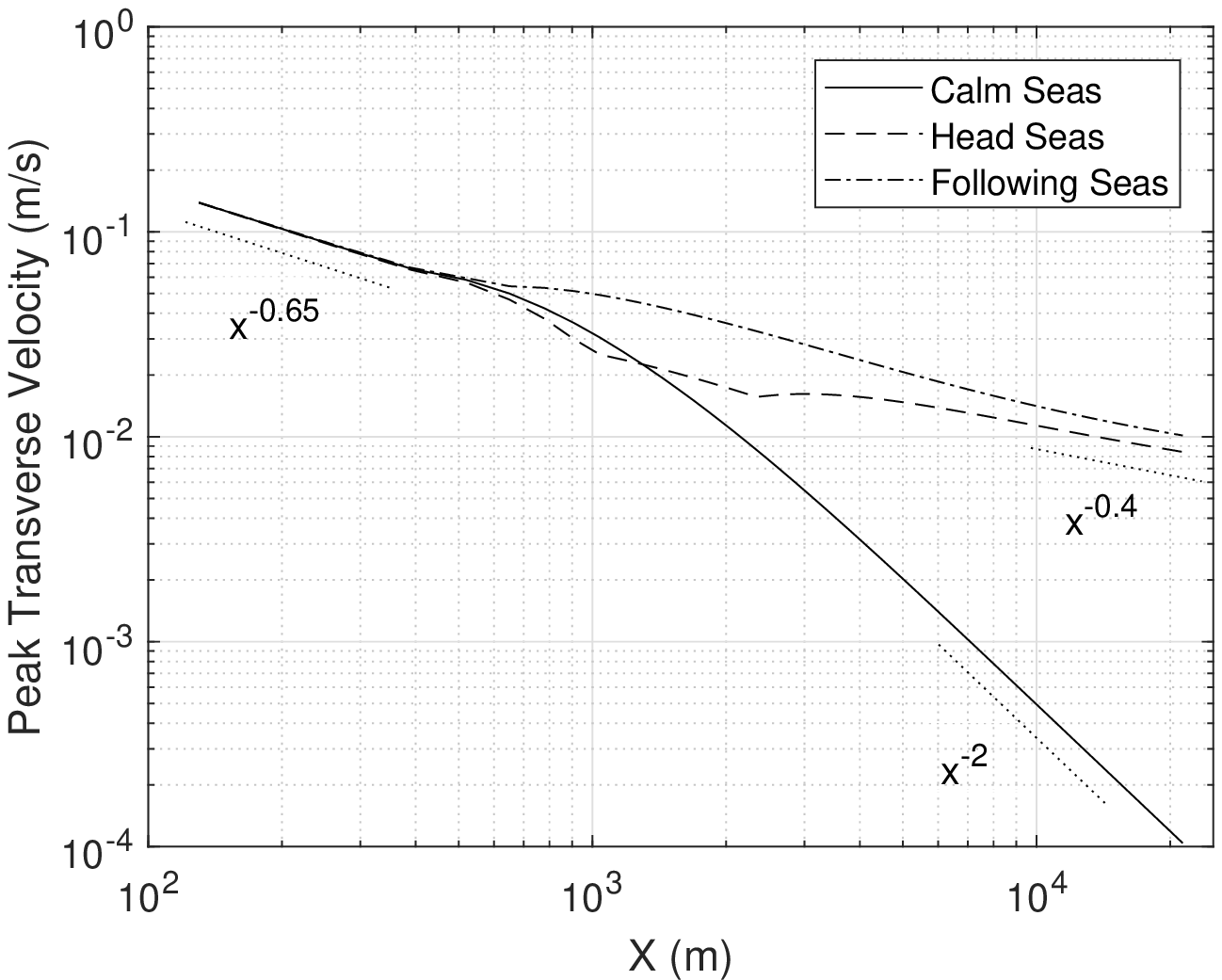}
\end{overpic}
\caption{Decay of the peak transverse velocity in calm, head, and following seas at Fr=0.35.  Surface waves with $\lambda = 10m$ and $a_s = 0.25m$.}
\label{fig:TransVEqn}
\end{figure}

The persistence of LTC can be demonstrated by tracking the peak transverse velocity in head seas and calm seas cases.  The peak transverse velocity ($\sqrt{v^2+w^2}$) is tangential to the rotational streamlines generated by the propeller or LTC.  Figure \ref{fig:TransVEqn} shows that both cases are dominated by propeller swirl in the near/far field (x \textless 1000m), but that LTC are several orders of magnitude stronger in the persistent wake.  In the near field, the propeller swirl decays proportional to $x^{-0.65}$ while the two propeller wakes are distinct, figure \ref{fig:axialContoursCalm}a.  \cite{sir96} suggest the swirl in an axisymmetric momentumless wake decays as $x^{-0.6}$ for cases with weak swirl and $x^{-0.75}$ for cases with strong swirl.  The swirling jets here are not in an axisymmetric momentumless wake, but this condition is considered more appropriate for comparison in the near field than a pure swirling jet.  Once the two propeller jets have merged (x$\approx$1000m) and the drag current has bifurcated, the transverse velocity decays proportional to $x^{-2}$.  This condition is more appropriate for comparison to a pure swirling jet and shows good agreement with the studies of \cite{shi08} and \cite{ewi99} based on conservation of angular momentum and equilibrium similarity.  

In the head seas case, a complex interaction between the decaying propeller swirl and growing LTC occurs between x=1000m and 2000m.  Around x=1000m, the propeller swirl has decayed below the peak transverse velocity of the outer circulations in figure \ref{fig:streamfunctionContoursHead}a.
As the inner circulations continue to grow, the outer circulations decay, until the inner circulations have a greater peak transverse velocity around x=2300m.
The inner circulations decay into the persistent wake with the peak transverse velocity proportional to $x^{-0.4}$.
Here, we define the beginning of the persistent wake as the point where the peak LTC induced transverse velocity is an order of magnitude greater than the peak propeller induced transverse velocity.  In head seas, this is occurs around x=6000m.

\begin{figure}
\centering
    \begin{overpic}[width=0.48\textwidth,tics=10]{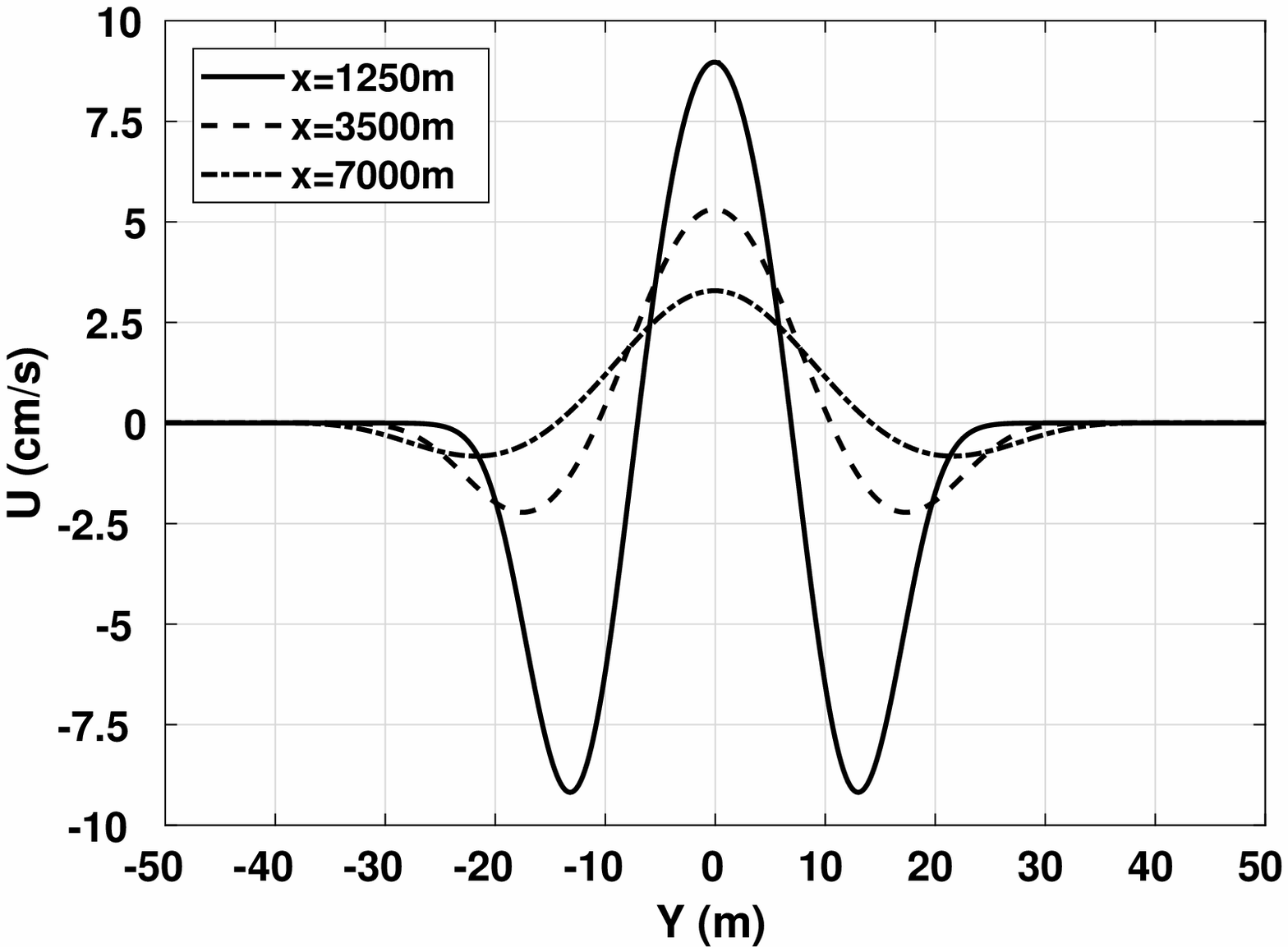}
	 \put (90,66) {\large$\displaystyle (a)$}
	\end{overpic} 
    \begin{overpic}[width=0.5\textwidth,tics=10]{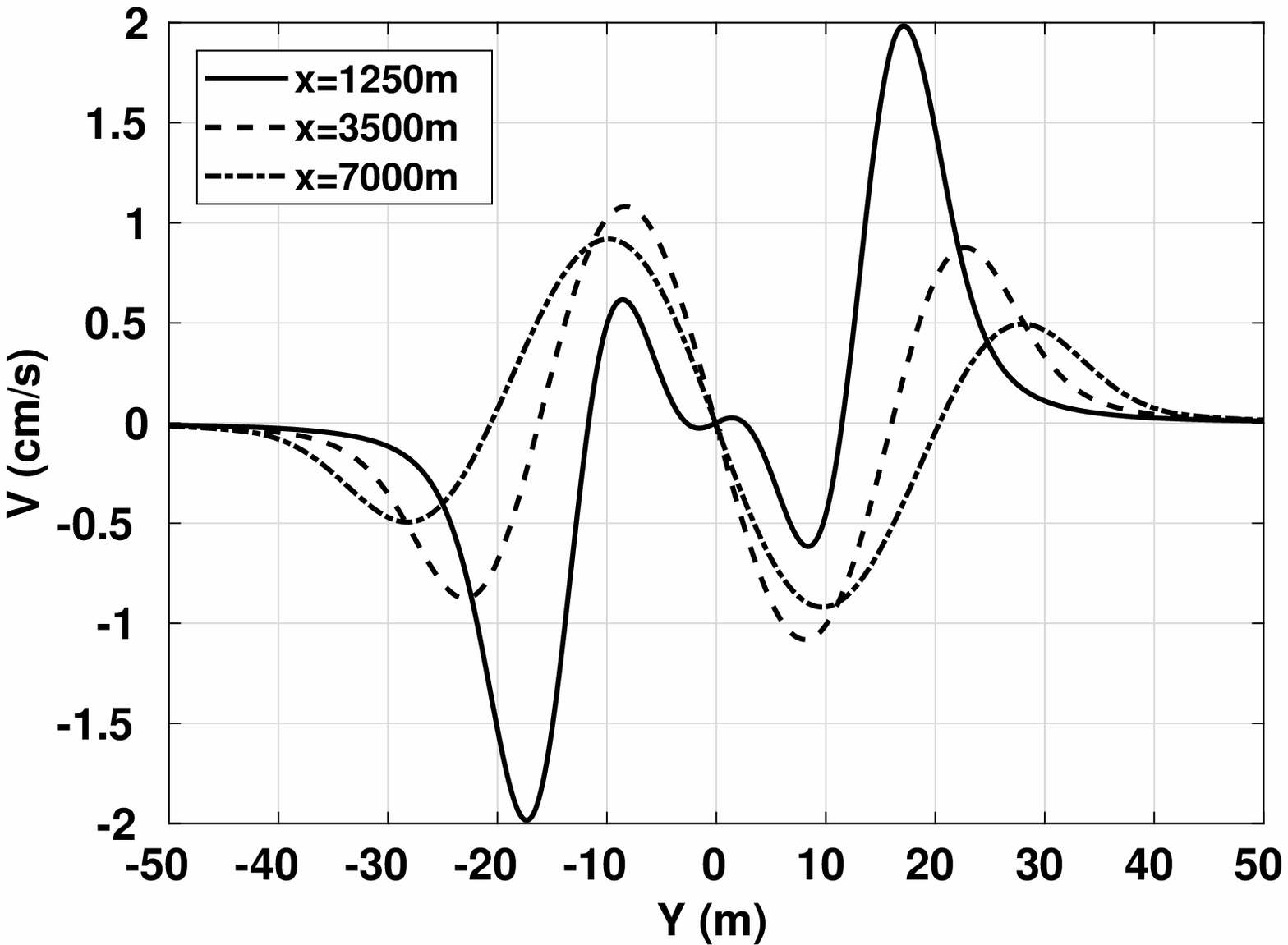}
	 \put (83,64) {\large$\displaystyle (b)$}
	\end{overpic} 
    \caption{Evolution of surface currents in head seas at $Fr=0.35$.  Surface waves with $\lambda=\unit[10]{m}$ and $a_s=\unit[0.25]{m}$. a) axial velocity b) transverse velocity.}
	\label{fig:currentProfilesHead}
\end{figure}
\begin{figure}
	\centering
     \begin{overpic}[height=6.5cm,tics=10]{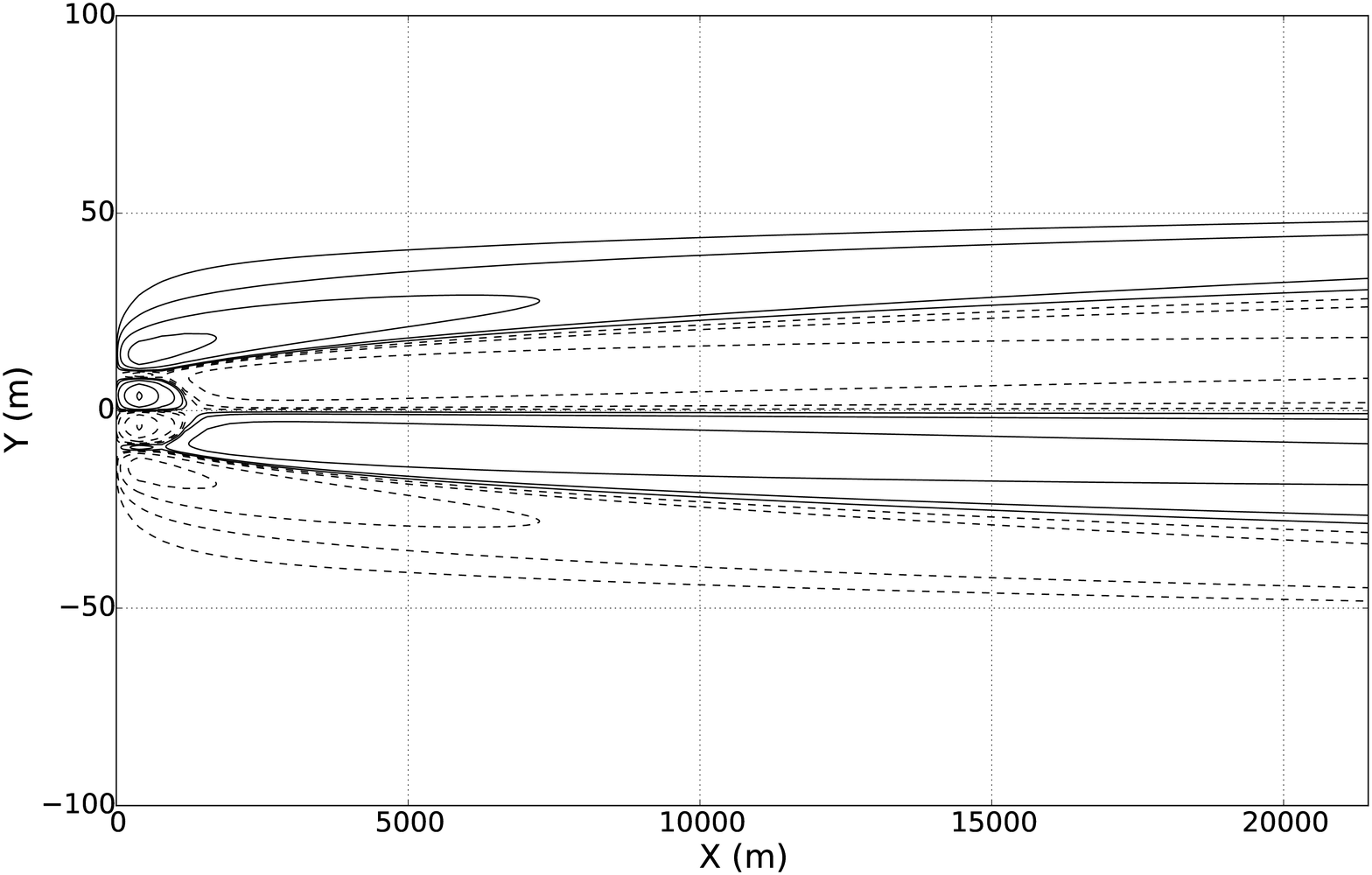}
     \put (65,55) {\vector(-0.5,-1.4){2.8}}
     \put (65,55) {\vector(0.0,-1){20.75}}
     \put (65,55) {\vector(0.5,-1.4){12.25}}
     \put (58,57) {\colorbox{white}{\parbox{0.1\linewidth}{Converging Zones}}}  
     \put (38,15) {\vector(-0.5,1.4){5}}
     \put (38,15) {\vector(0.5,1.4){9.1}}
     \put (33,11) {\colorbox{white}{\parbox{0.1\linewidth}{Diverging Zones}}} 
     \end{overpic} 
    \caption{Contour map of transverse--component of velocity on ocean surface, in head seas at $Fr=0.35$.  Surface waves with $\lambda=\unit[10]{m}$ and $a_s=\unit[0.25]{m}$. Contours = +/- 0.05, 0.15, 0.5, 1.5, 3.5, 5 cm/s.  Dashed lines are negative contours.}
    \label{fig:currentContoursHead}
\end{figure}

Figure \ref{fig:currentProfilesHead} presents the evolution of axial and transverse components of the surface velocity in head seas as a function of the transverse coordinate  $y$.  Plots are presented in figure \ref{fig:currentProfilesHead} for the same three distances behind the ship as shown in figure \ref{fig:currentProfilesCalm}.  Evolution of axial velocity for the case of head seas shown in figure \ref{fig:currentProfilesHead}a is similar to the behavior observed in the case without surface waves, as depicted in figure \ref{fig:currentProfilesCalm}.   Several kilometers behind the ship only a small residual of the axial velocity remains.  However, the evolution of the  transverse velocity in the presence of ambient surface waves is totally different than in the absence of a surface wave field.   Its structure reflects the four LTC developed due to interaction of the axial current with surface waves.  The magnitude of the LTC--related transverse surface current grows during the initial stage of LTC development  and then stays relatively unchanged (figure \ref{fig:currentProfilesHead}b).  The magnitude of the flow caused by the pair of inner LTC changes very little as the distance increases from \unit[3500]{m} to \unit[7000]{m}.  The magnitude of the transverse flow related to the outer pair of LTC does decrease slowly as the width of these circulations increases.  The LTC persist long ($O(10km)$) after the circulatory flows caused by the propellers disappear, as seen in the distribution of transverse surface velocity on the ($x$,$y$) plane in figure \ref{fig:currentContoursHead}.  The role of ambient surface waves and generation of LTC in forming the persistent wake is especially noticeable when figure \ref{fig:currentContoursHead} is compared to figure \ref{fig:currentContoursCalm}, which corresponds to the calm--seas case.  The persistence of the LTC generated transverse velocity is due in part to the vortex force continuing to reinforce the LTC while the axial current exists and  is in part due to the reduced level of turbulence in the region of the persistent wake.   Before the LTC develop, the transverse component of the surface flow is dominated by propeller--generated circulatory flows that, for this ship, rotate outward (figure \ref{fig:currentContoursCalm}).  The transition from the region of dominance of propeller--induced flow to the region of the persistent wake is reflected in figure \ref{fig:currentContoursHead} by the centerline saddle--point region observed about \unit[1000]{m} behind the ship.  
 
For the head seas case, LTC that form the persistent wake produce a strong convergence zone along the center of the wake due to inward--rotating inner circulations.  Also, there are two divergence zones produced between the inner-- and outer--rotating circulations.  Additionally, there are two so--called weak convergence zones at the outer edges of the wake, which are produced by the two outer, outward--rotating  circulations.  While the transverse surface velocity induced by the LTC is relatively low and cannot strongly affect the short surface waves directly, the existence of  persistent regions of convergence and divergence can cause significant redistribution of the SAS films, which in turn will modify  the short surface waves.  In the case of head seas, the convergence zone near the centerline will create a region of high SAS concentration, which will reduce the amplitude of the short surface waves  and, correspondingly, produce a dark centerline wake in SAR imagery.  In addition, SAS concentration will also occur along the outer weak convergence zones, but the strength of these outer convergence zones is clearly weaker than the centerline convergence zone.  Depending on the strength of the outer LTC, these outer convergence zones can form an additional set of streaks in SAR imagery.  

\subsection{Wake evolution for ship in following seas}
\begin{figure}
	\centering
\begin{overpic}[width=0.75\textwidth,tics=10]{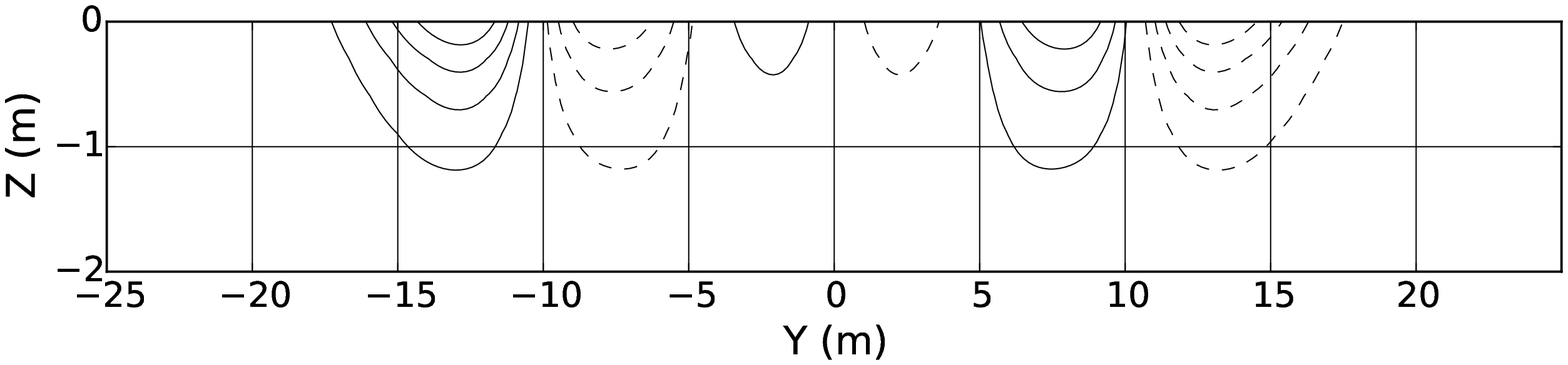}
\put (10,18) {\large$\displaystyle (a)$}
\end{overpic}
\begin{overpic}[width=0.75\textwidth,tics=10]{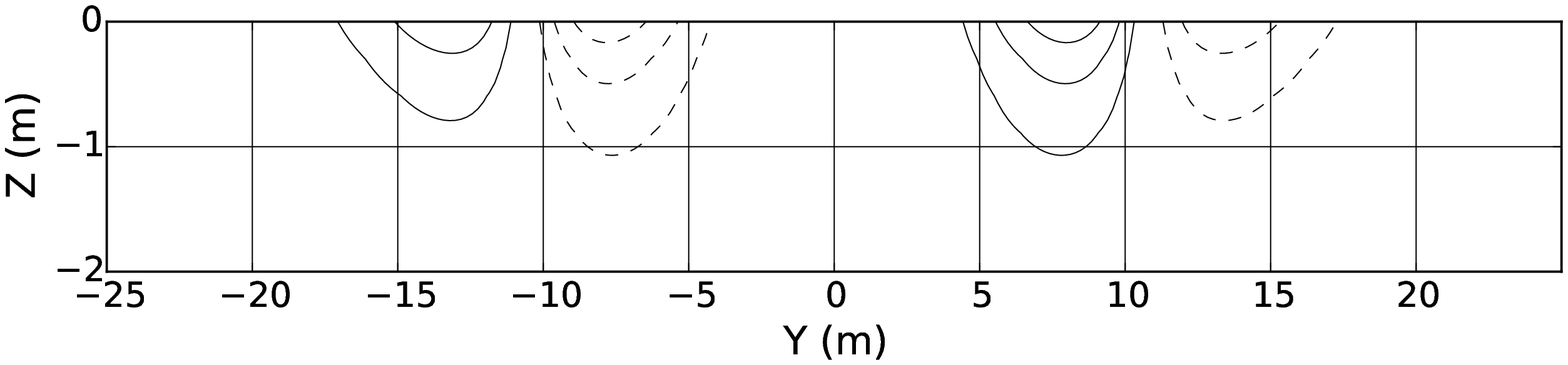}
\put (10,18) {\large$\displaystyle (b)$}
\end{overpic} 
\begin{overpic}[width=0.75\textwidth,tics=10]{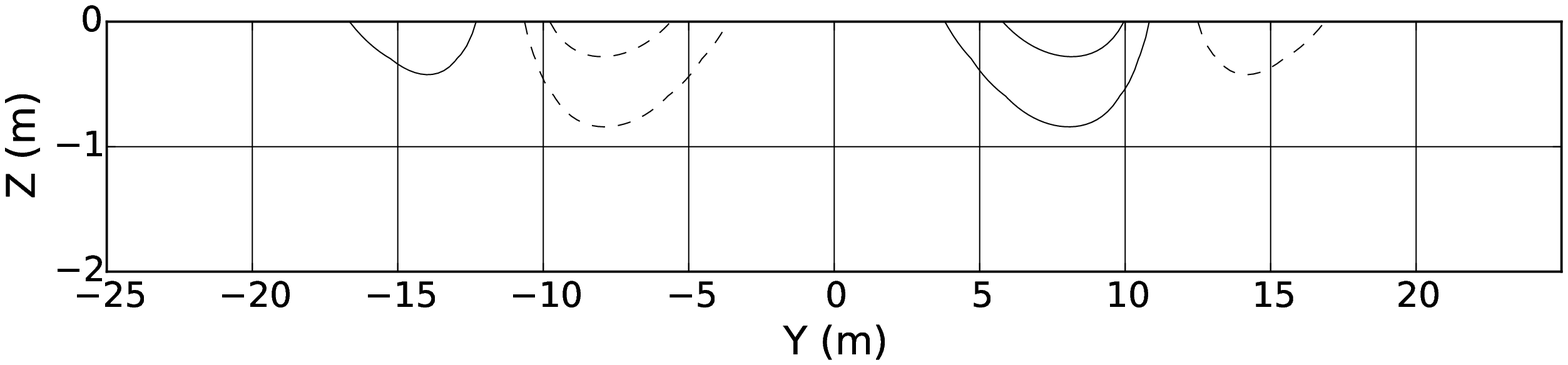}
\put (10,18) {\large$\displaystyle (c)$}
\end{overpic}
\caption{Contour maps of the horizontal component of the vortex force $F_y$ at (a) $x$ = 450m, (b) $x =$ 850m, and (c) $x =$ 1250m in following seas at Fr = 0.35.  Surface waves with $\lambda = 10m$ and $a_s = 0.25m$.  Contour spacing = 0.001 N/kg.  Peak vortex force at x=450m is 0.0049 N/kg.  Dashed lines are negative contours.}
\label{fig:VF_YFS}
\end{figure}

\begin{figure}
\centering
\begin{overpic}[width=0.75\textwidth,tics=10]{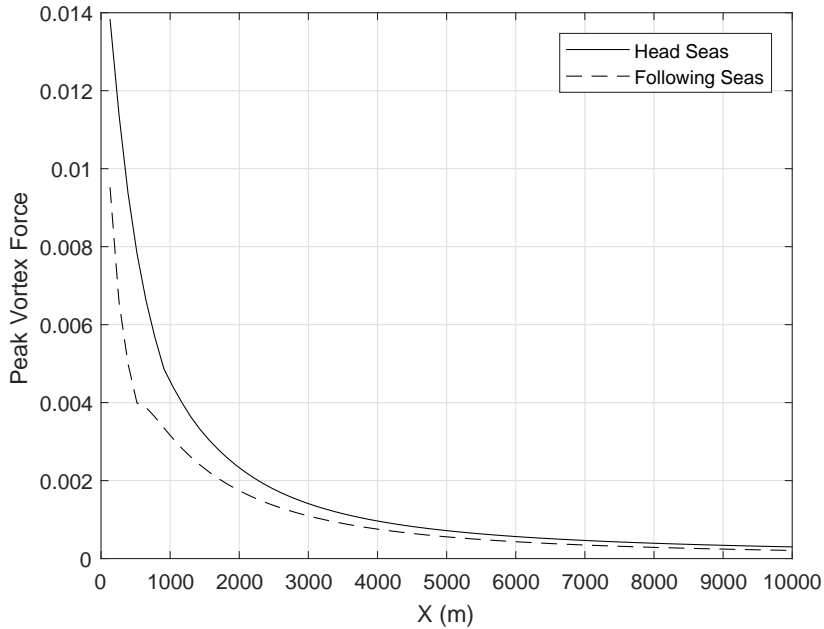}
\end{overpic}
\caption{Decay of the peak horizontal component of vortex force (N/kg) in head and following seas at Fr=0.35.  Surface waves with $\lambda = 10m$ and $a_s = 0.25m$.}
\label{fig:VF_Decay}
\end{figure}

In comparison to the case of head seas, the horizontal component of the vortex force in following seas reverses sign, figure \ref{fig:VF_YFS}, but shows a similar magnitude and shape with four regions with alternating sign.  These regions gradually spread, but at a slower rate than in the case of head seas.  The outermost regions act inwards towards the centerline, thus reducing the rate of spreading in comparison to the case of head seas where the current is pushed further outboard by the outboard--acting forces.  The vortex force in following seas also dissipates sooner than in head seas, figure \ref{fig:VF_Decay}.  

\begin{figure}
	\centering
        \begin{overpic}[width=0.75\textwidth,tics=10]{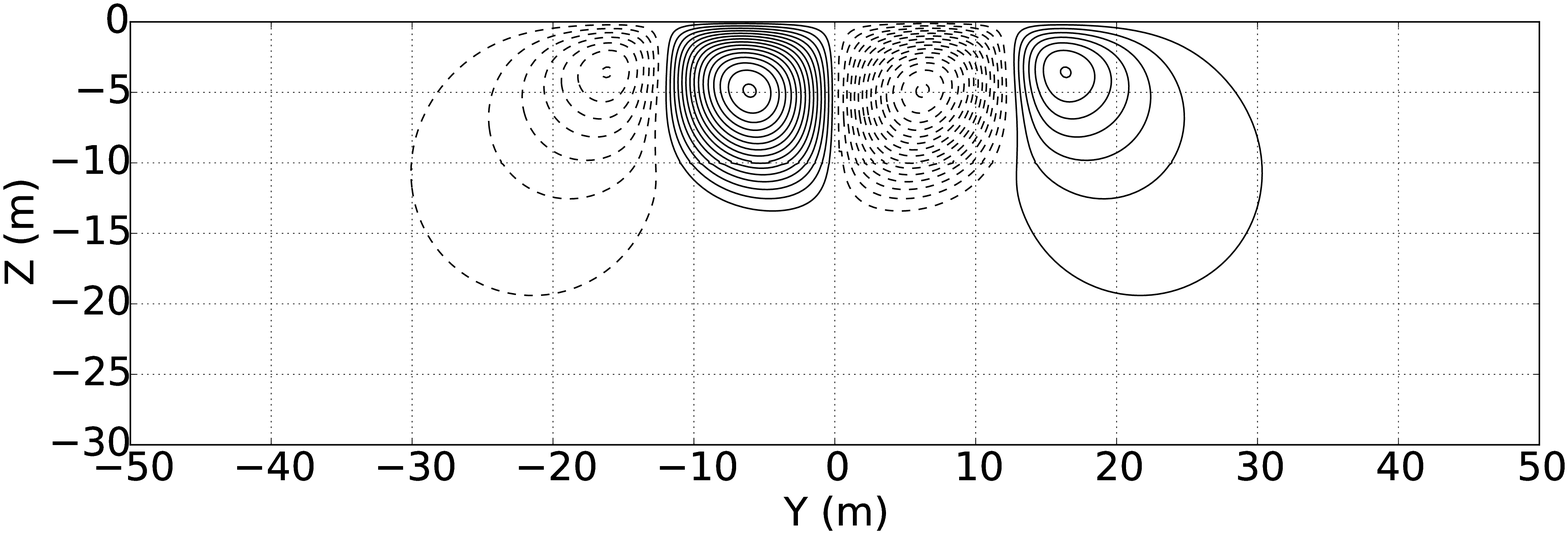}
         \put (11,30) {\large$\displaystyle (a)$}
	\end{overpic} 
          \begin{overpic}[width=0.75\textwidth,tics=10]{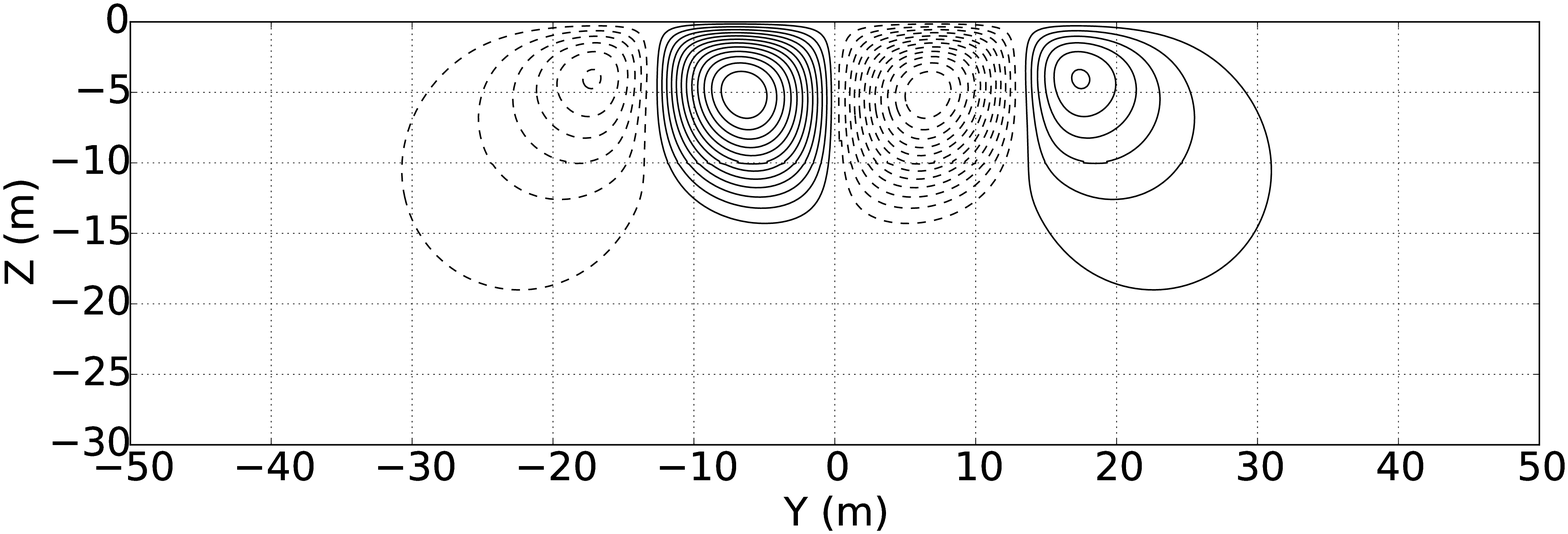}
	 \put (11,30) {\large$\displaystyle (b)$}
	\end{overpic} 
            \begin{overpic}[width=0.75\textwidth,tics=10]{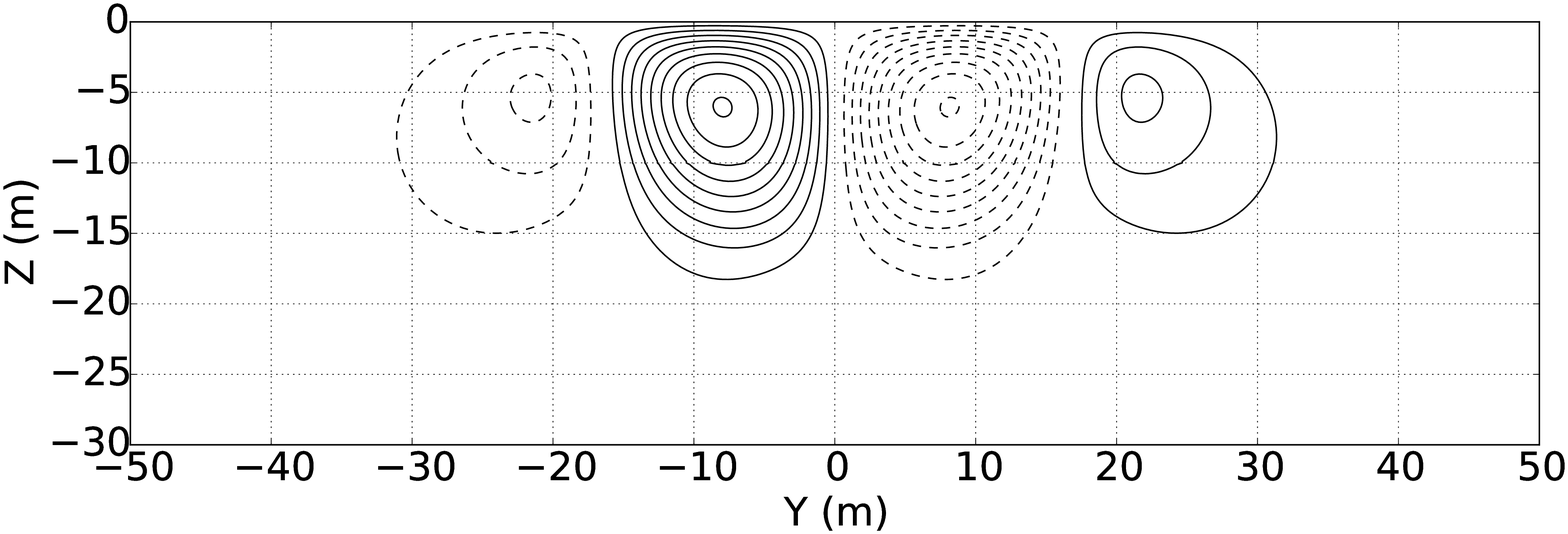}
	 \put (11,30) {\large$\displaystyle (c)$}
	\end{overpic} 
	
    \caption{Contour maps of streamfunction at (a) $x = \unit[1250]{m}$,  (b) $x =$ \unit[1650]{m}, and (c) $x =$ \unit[3500]{m}, in following seas at $Fr=0.35$.  Surface waves with $\lambda=\unit[10]{m}$ and $a_s=\unit[0.25]{m}$.  Contour spacing = $\unit[0.005]{m^2/s}$.  Propeller swirl and inner LTC are reinforcing in following seas.}
    \label{fig:streamfunctionContoursFollowing}
\end{figure}

The development of the transverse flow in the persistent wake for the case of following seas is shown in figure \ref{fig:streamfunctionContoursFollowing}  for the same distances behind the ship as presented in figure  \ref{fig:streamfunctionContoursHead} for the case of head seas.  Parameters of the ambient surface waves are taken to be the same as in the previous case:  sinusoidal wave with amplitude of \unit[0.25]{m} and wavelength of \unit[10]{m}.  Here the development of four LTC with structure similar to the case of head seas can be observed, but with an important difference: the directions of the circulations are the opposite to the direction of the circulations in the head seas case.  Specifically, the pair of inner circulations rotate outward and outer pair rotate inward.  Since the initial fluid motions induced by the two propellers are outward rotating, they quickly merge with the inner LTC, unlike the propeller--induced motions in the head seas case.  

The inner LTC in following seas grow faster in magnitude and size than in head seas.  The inner LTC generate greater transverse velocities than the decaying propeller swirl by x=500m.
Despite the lower magnitude vortex forces, the case of following seas generates greater transverse velocities than the head seas case for the range of study, figure \ref{fig:TransVEqn} due to these larger inner circulations.  The LTC continue to grow in strength for several hundred meters beyond where they surpass the propeller swirl, before asymptotically approaching an equal decay rate to the head seas case with the transverse velocity in the persistent wake again proportional to $x^{-0.4}$.  In following seas, the persistent wake is observed to begin around x=5000m, 1000m sooner than the head seas case. 

\begin{figure}
\centering
        \begin{overpic}[width=0.495\textwidth,tics=10]{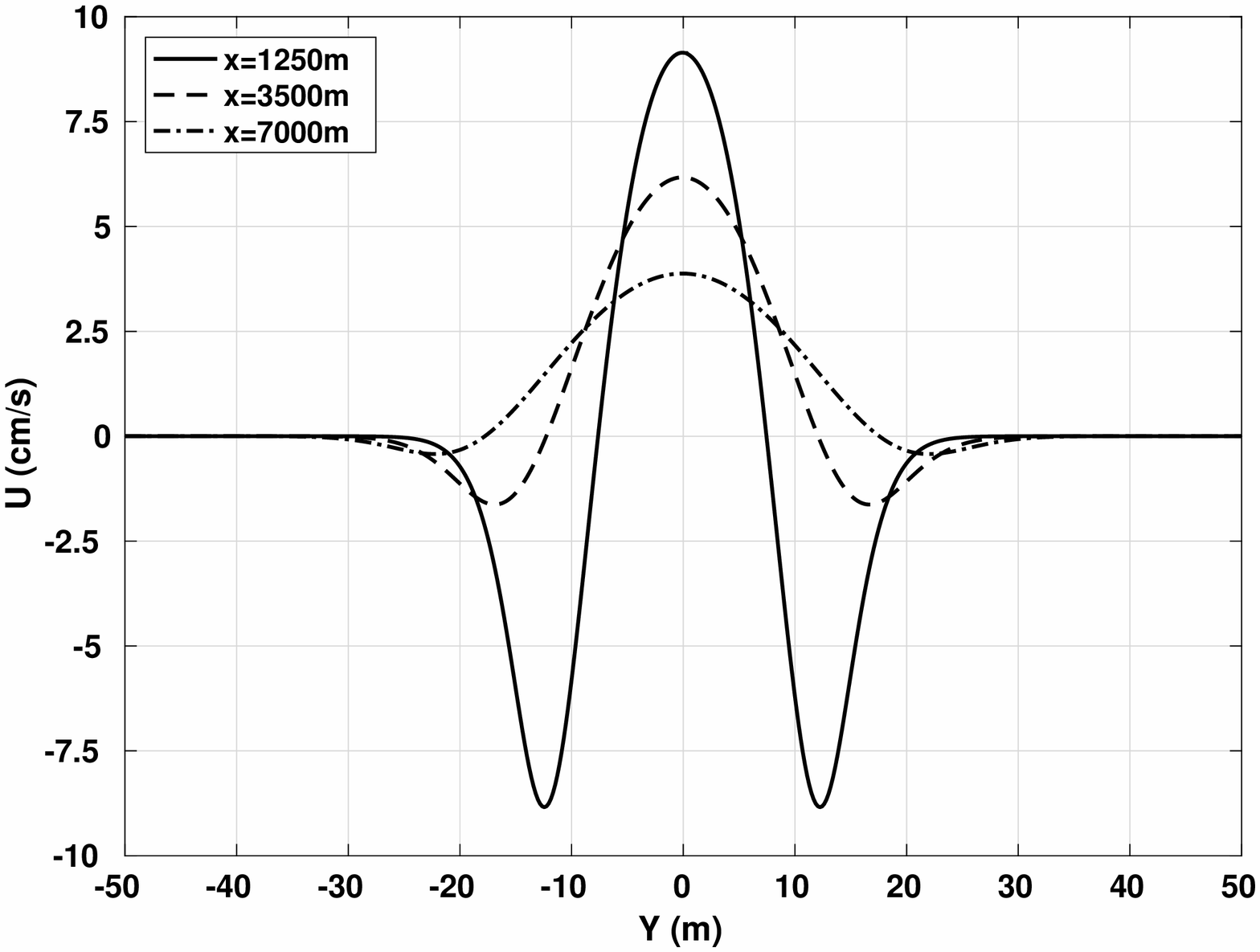}
	 \put (90,68) {\large$\displaystyle (a)$}
	\end{overpic} 
	        \begin{overpic}[width=0.48\textwidth,tics=10]{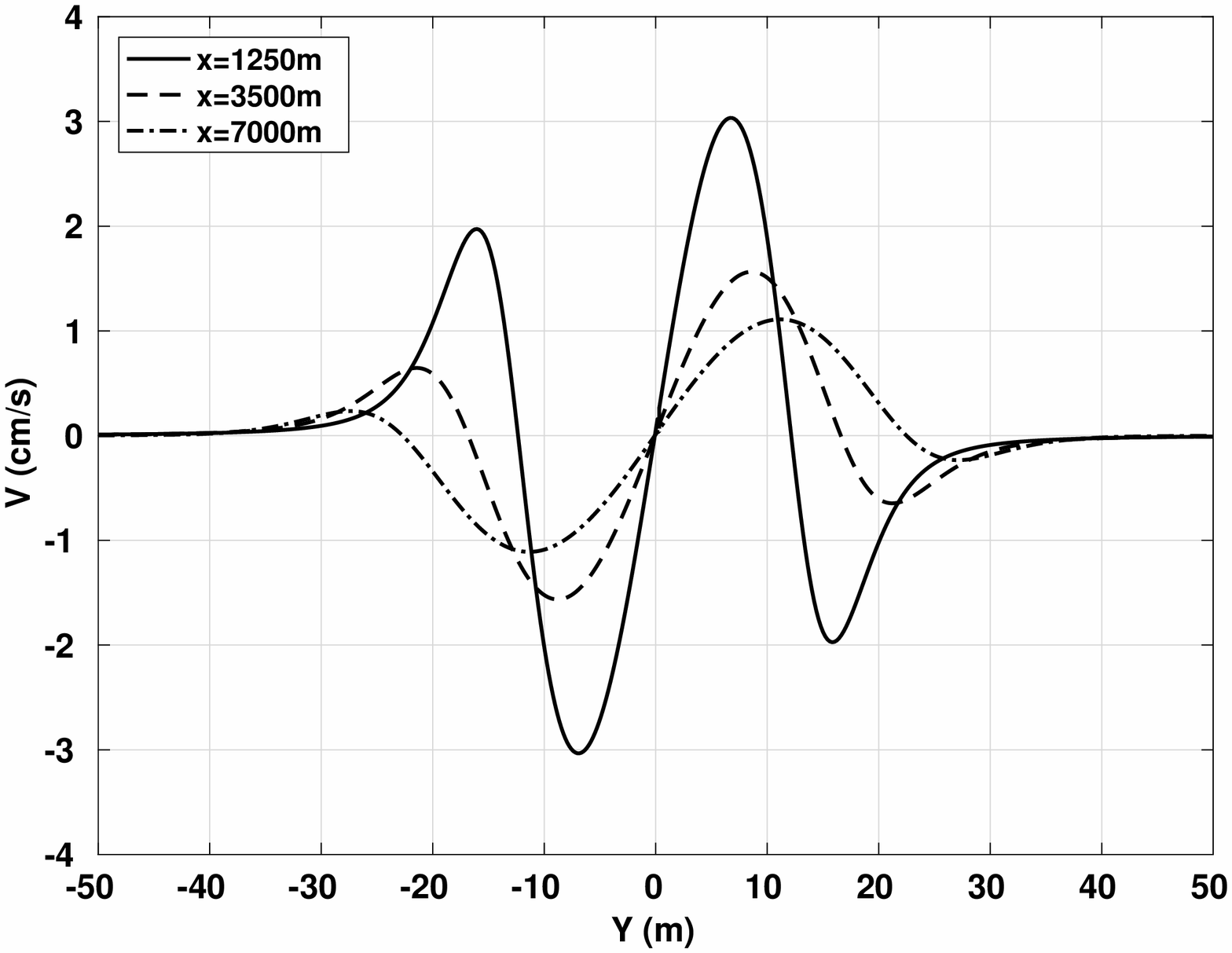}
	 \put (90,70) {\large$\displaystyle (b)$}
	\end{overpic} 
        \caption{Evolution of surface currents in following seas at $Fr=0.35$.  Surface waves with $\lambda=\unit[10]{m}$ and $a_s=\unit[0.25]{m}$. a) axial--component of velocity b) transverse--component of  velocity.}
	\label{fig:currentProfilesFollowing}
\end{figure}

\begin{figure}
	\centering
    \begin{overpic}[height=6.5cm,tics=10]{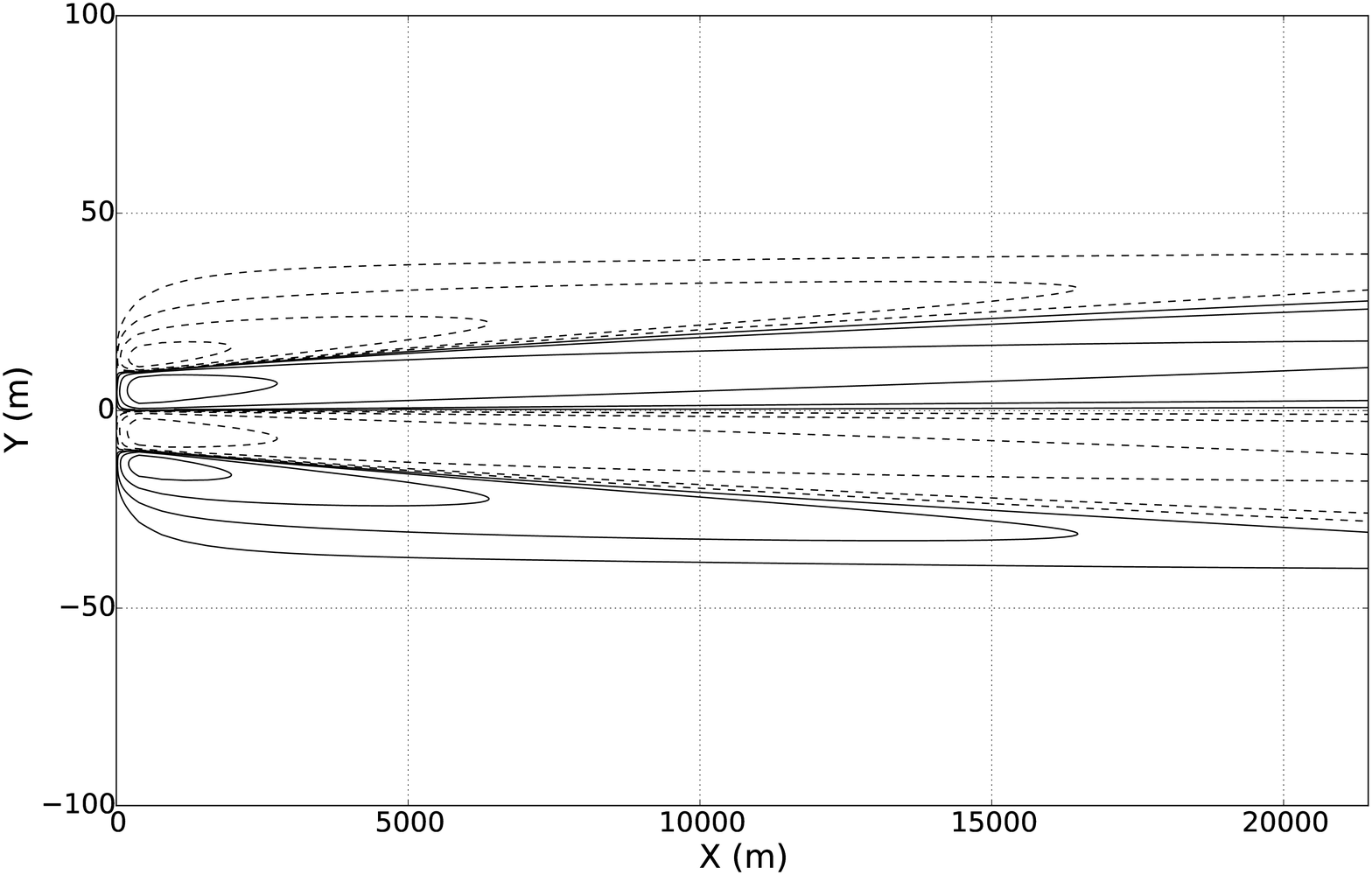}
    \put (65,55) {\vector(-0.25,-1.4){2.6}}
     \put (65,55) {\vector(0.25,-1.4){4.9}}
     \put (58,57) {\colorbox{white}{\parbox{0.1\linewidth}{Converging Zones}}}   
     \put (39,15) {\vector(-0.5,1.4){3}}
     \put (39,15) {\vector(0.0,1){19.5}}
     \put (39,15) {\vector(0.5,1.4){11}}
     \put (33,11) {\colorbox{white}{\parbox{0.1\linewidth}{Diverging Zones}}}
     \end{overpic}
	\caption{Contour map of transverse--component of velocity on ocean surface, in following seas at $Fr=0.35$.  Surface waves with $\lambda=\unit[10]{m}$ and $a_s=\unit[0.25]{m}$.  Contours = +/- 0.05, 0.15, 0.5, 1.5, 3.5, 5 cm/s.  Dashed lines are negative contours.}
    \label{fig:currentContoursFollowing}
\end{figure}

The surface velocities in the ship wake in following seas are presented in figure \ref{fig:currentProfilesFollowing}.  Figure \ref{fig:currentProfilesFollowing}a shows the evolution of the axial velocity, which is fairly similar to its behavior in the previous cases of calm seas (figure \ref{fig:currentProfilesCalm}a) and head seas (figure \ref{fig:currentProfilesHead}a).  The presence of the upwelling zone along the centerline is observed to increase the persistence of the surface centerline axial velocity while the outboard surface drag current is suppressed by the downwelling zones.  These differences in the  near--surface axial velocity contribute to the stronger inner LTC and weaker outer LTC observed in figure \ref{fig:streamfunctionContoursFollowing}, compared to the case of head seas (figure \ref{fig:streamfunctionContoursHead}).  In following seas, the transverse velocity at the surface (figure \ref{fig:currentProfilesFollowing}b) has a direction opposite to the direction of the velocity in the head seas case (figure \ref{fig:currentProfilesHead}b).  The LTC extend well beyond the far wake as shown in figure \ref{fig:currentContoursFollowing}, which presents the distribution of the transverse surface velocity on the $(x,y)$ plane.  Because of the change in the sign of the circulations, LTC in the persistent wake for the case of following seas create a divergence zone along the centerline of the wake, two strong convergence zones on each side of the centerline and two weak divergence zones at the outer edges of the wake.  The distribution of the SAS films due to these LTC will be different from the case of head seas.  The two strong convergence zones offset from the centerline will create bands with high SAS concentration and, correspondingly, dark streaks in the SAR images, producing a railroad--track wake.  

\subsection{Effect of propeller rotation direction and propeller count}
From the previous discussion, it is obvious that the exact configuration of the ship and the propellers are not the main factors in the formation of the persistent wake. \cite{pel93} reported similarity in wake evolution between a twin--propeller destroyer and a single--propeller frigate.   Thus, the effects of propeller--rotation direction and a change in the number of propellers were considered.  The results of the study will be presented in a separate paper.  Here, we note that the simulations show that the structure of the persistent wake generated by same ship model (twin propellers) does not appreciably change when the direction of propeller rotation is changed.

To examine the effect of the propeller configuration on the structure of the persistent wake, the IDP of the present model was changed from a twin--propeller configuration to a single--propeller configuration, and simulations were run for the head seas condition.  The single propeller was sized to have an equal loading to the previously modeled twin propeller configuration.      

\begin{figure}
	\centering
    \begin{overpic}[height=6.5cm,tics=10]{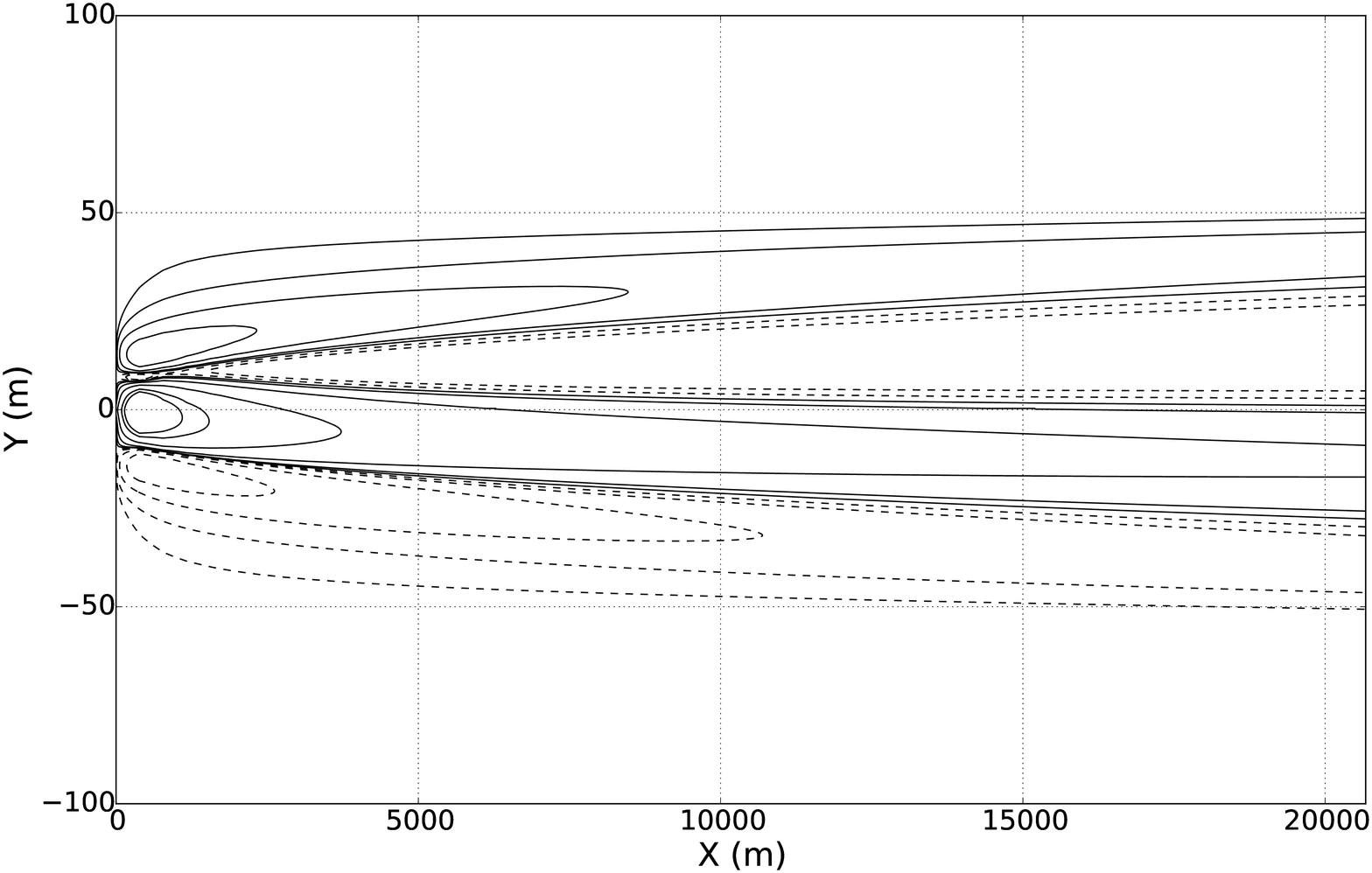}
      \put (65,55) {\vector(-0.5,-1.4){2.8}}
     \put (65,55) {\vector(0.0,-1){20.5}}
     \put (65,55) {\vector(0.5,-1.4){12.25}}
     \put (58,57) {\colorbox{white}{\parbox{0.1\linewidth}{Converging Zones}}} 
     \put (38,15) {\vector(-0.5,1.4){5}}
     \put (38,15) {\vector(0.5,1.4){9.1}}
     \put (33,11) {\colorbox{white}{\parbox{0.1\linewidth}{Diverging Zones}}}
     \end{overpic}
    \caption{Contour map of the transverse--component of velocity on ocean surface, in head seas at $Fr=0.35$ with a single propeller.  Surface waves with $\lambda=\unit[10]{m}$ and $a_s=\unit[0.25]{m}$.  Contours = +/- 0.05, 0.15, 0.5, 1.5, 3.5, 5 cm/s.  Dashed lines are negative contours.}
    \label{fig:currentContoursSingleProp}
\end{figure}

The contours of the transverse component of surface velocity for the single propeller on the ($x$,$y$) plane is shown in figure \ref{fig:currentContoursSingleProp}.  This distribution is different from the two-propeller configuration shown in figure \ref{fig:currentContoursHead} as the centerline convergence zone is initially offset due to the circulatory flow produced by the single propeller.  However, as the distance behind the ship increases and initial propeller--induced circulatory flow decays, the developing LTC start to define the structure of the persistent wake and the distributions of transverse surface velocity in figures \ref{fig:currentContoursSingleProp} and \ref{fig:currentContoursHead} (single versus twin propellers, both head seas) become similar, but with reduced symmetry in the single--propeller case.  This result is to be expected as the LTC are driven by the axial velocity profile, which behaves as a single jet in the twin propeller configuration, once the two thrust currents merge (figure \ref{fig:axialContoursCalm}).  

\section{Comparison with experimental data}
A rigorous comparison of the theoretical insights with available experimental SAR data is neither feasible nor in scope of the analysis reported herein.  On the one hand, such a comparison will require an advanced model that describes the dynamics of SAS films under the influence of LTC and surface waves.  That model should include a description of all the relevant factors that affect surface films, including diffusion, degradation and scavenging.  The formulation of such a model is currently in progress.  On the other hand, the experimental data that contains detailed measurements of parameters of persistent wakes is very limited.  For example, as was mentioned previously, \cite{erm14} detected the presence of circulations by registering transverse velocities on the order of centimeters per second far behind a passing ship.  However, the salient parameters of the axial current and of surface waves were not reported.  There are many papers that provide SAR images of ship wakes (see, for example, \cite{top11,mun87,sol10,lyd88}), but often sea conditions  and ship parameters at the time the images were obtained are not well documented, making it difficult to compare these experimental results with the present theory. 

The best documented series of full-scale experiments, to which we have access, was conducted during the 1989 Office of Naval Research Field Experiment \citep{pel93}.  During these experiments, the wakes behind both a U.S. Navy destroyer and a U.S. Navy frigate were investigated.  Specifically, {\textit{in situ}} measurements of the ocean surface tension were collected to coincide with the collection of airborne SAR imagery of the ocean surface.  The wind and surface wave conditions were also recorded.  The environmental conditions for the experiments with the Navy frigate were very different from the environmental conditions during the experiments with the Navy destroyer.  For the frigate experiments, the reported wind was extremely light and the sea surface was totally still.  Under such calm weather conditions, the initial hull-- and propeller--generated perturbations to the SAS film can persist for a relatively long time, thus producing a long--lasting surface disturbance, which can be observed only under the described conditions.  Perturbations to the surface tension along the  wake edges of the frigate in calm seas were measured to be less than those along the wake edges of the destroyer operating under conditions with ambient surface waves.  These measured differences are qualitatively in agreement with the prediction of our simulation.

Most of the experimental data collected for the US Navy destroyer was obtained in conditions of following seas.  \textit{In situ} measurements of surface tension across the ship wake were taken from a small boat that crossed the wake.  Several bands with a high concentration of SAS films were registered.  Not all bands registered with the \textit{in situ}  measurements can be clearly identified, since not each of them can be seen in the accompanying SAR imagery (see figure \ref{fig:FollowingSeaSAR}).  Nevertheless, the \textit{in situ} measurements do confirm that the SAS concentration is low along the centerline of the wake and that there are two bands, one on each side of the wake centerline, where SAS concentration is high.  These two bands correlate with the position of the two dark streaks that are seen in figure \ref{fig:FollowingSeaSAR}.  The length of the wake in the image is about \unit[7]{km}.   Based on the description of figure \ref{fig:FollowingSeaSAR} by \cite{mil93}, the distance between the dark streaks (railroad tracks) in the image is roughly estimated to be about \unit[70]{m}.  The formation of such a railroad--track wake for the case of following seas has been described in this paper.  The transverse surface current generated in the simulation of the case of following seas (figure \ref{fig:currentContoursFollowing}) shows the distance between converging zones is about \unit[50]{m}.  This value is in reasonably good agreement with the estimate obtained from the SAR image.  The difference can be caused by many factors such as an imperfect model of initial current at the IDP, uncertainty in the ambient surface--wave parameters, or the potential effect of ocean stratification on the development of LTC and other factors. 

It also should be noted that the railroad--track structure of the persistent wake in the SAR image begins only beyond the distance of about \unit[3]{km} behind the ship.  At shorter distances, the wake appears as a centerline wake.  Such behavior is observed in the region of the far wake, where the ship--generated perturbations are still relatively strong and the turbulence has a more significant damping effect than does the SAS film on the amplitude of the short surface waves.  Once the turbulence decays, then the LTC--induced structure of the wake becomes more apparent.

Although the following seas case was the focus of the systematic study of the ONR set of experiments, a SAR image of the wake of the same US Navy destroyer in head seas was also obtained.  That image is presented in figure \ref{fig:HeadSeaSAR}.  Only a centerline wake is seen in the SAR image, which is in agreement with the predictions for the case of head seas made in  this paper.  The dark centerline streak is the manifestation of a strong convergence zone, which is produced by the LTC at the center of the wake (see figure  \ref{fig:currentContoursHead}).  Notice that initially the centerline wake is wide, which is caused by the presence of turbulence in the far wake region behind the ship.  As the far wake transforms into a persistent wake, which is defined by the evolving LTC, the centerline streak narrows. 

\section{Conclusions}

An extended region of a surface ship wake far behind the ship, where original perturbations produced by the motion of the hull and propellers have decayed significantly, has been identified and characterized.  The existence and configuration of this wake segment, which we refer to as the {\textit{persistent wake}}, is as much the result of ocean dynamics as of ship configuration and operating parameters.  
The persistent wake arises as a result of the interaction between ship--induced currents and an energetic surface--wave field, which produces large--scale, nearly--inviscid secondary circulatory flows orthogonal to the direction of the ship drag and thrust currents.  The interaction with the surface--wave field induces a so--called vortex force \citep{cra76} on ship--generated non--uniform currents, which then generates the circulations.  These circulations, first described by \cite{bas11}, are similar in character to the well--known Langmuir circulations \citep{lan38}; therefore, we identify them as Langmuir--type circulations (LTC).  Unlike naturally--generated Langmuir circulations, which arise as a result of an instability in the wind--driven surface currents, LTC are formed around a ship--induced non--uniform current system comprised of thrust currents (direction opposite to the ship motion) and drag currents (direction same as the ship motion).  The structure and magnitude of the resultant LTC depend on the magnitude and relative direction of the ship--induced currents and the ambient surface--wave field.   The description of the ship--induced currents follow the computational approach to establish an initial data plane (IDP) introduced by \cite{min88}. 

Numerical simulations reveal several important features of the development of a persistent wake.  First, the persistent wake does not develop in the absence of a surface--wave field or if the direction of the ship--induced currents and the propagation direction of the surface--wave field are nearly orthogonal. Second, the effects of the interaction of the ship currents and surface--wave field are not immediately apparent, since it takes time for LTC to develop.  Third, the structure of LTC and, correspondingly, of the persistent wake is different for the case with head seas (ship heads into the waves) than for the case with following seas (ship travels with the waves). This difference occurs because the sign of the vortex force depends on the direction of the surface waves relative to the direction of the ship--induced currents.  For the particular model of the ship studied in the present paper, the development of the large--diameter circulations can take minutes to form, and the magnitude of the transverse (cross--track) surface flows induced by the LTC are on the order of centimeters per second.   At five to eight kilometers downstream of the IDP, the cross--track extent of the system of circulations can be three to four times the width of the beam of the ship.  The resulting system of circulations can exist for tens of kilometers behind the ship, which is consistent with their large size and the relatively low eddy viscosity of the background fluid. 
 
Previous experiments \citep{pel93, erm14} confirm that the appearance of the persistent wake is defined in remote SAR and optical imagery by the change in spectral density of the short surface waves, which is a result of the redistribution of SAS films by the LTC--induced surface flow in the wake.  In the case of head seas, the LTC--induced flow at the surface creates a strong convergence zone at the center of the wake.  The concentration of SAS films in this region is increased, thereby forming a centerline wake and producing a dark streak in SAR imagery.  In the case of following seas, the LTC--induced flow at the surface creates two convergence zones on each side of the centerline, several tens of meters apart.  Due to an increase in the concentration of the SAS film at these zones, dark streaks are produced in the SAR images, which appear as so--called railroad--track ship wakes.  These simulation results are in good agreement with the experiment--based observations presented by \cite{pel93}.

There are some other possible LTC related effects of potential importance for understanding the physics of the wake.  For example, developing LTC can strongly affect the lifetime of bubble clouds created by the ship propellers, since LTC can create converging zones (downwelling) or diverging zones (upwelling) at the centerline of the wake for head and following seas, respectively. The lifetime of the bubble clouds depends on the time for the bubbles to rise to the surface.  LTC-generated flow will inhibit the upward motion of the bubble in head seas and contribute to this motion in following seas.  Thus, the lifetime of the bubble clouds can be expected to be noticeably shorter in the case of following seas, for ship configurations similar to the one studied here.  

The present paper is the first step in the study of the role of LTC in the physics of ship wakes.  The need for further work is recognized.  For example, the effects of ship--propeller configuration and the direction of their rotation should be studied.  Also, a comprehensive model of SAS film--induced effects is required.  It should include the mechanisms of film redistribution by surface flows and turbulence, film degradation and diffusion, the effect of scavenging of SAS on film concentration, as well as the effects of bubbles and other contaminants on these processes.  Present models of the persistent wake can be measurably improved by better describing the effects of SAS films and near--surface turbulence on the spectrum of short surface waves, which are responsible for the manifestation of ship--induced anomalies in sea--surface imagery.       
 
\textit{Acknowledgements}. The authors are thankful to the President of Cortana Corporation, Mr. K.J. Moore, for continuing support and help in improvement of the paper.  The authors also thank Dr. John G. Pierce and Dr. Rodney D. Peltzer for their insight and fruitful discussions. 

\bibliographystyle{plainnat}
\bibliography{PersistenceOfShipWakes}

\appendix
\section{Theoretical model for a surface--combatant wake}

An empirical--analytical representation of the wake one-half ship length downstream, figure \ref{fig:IDP}, is established based on a formulation similar to \cite{min88}.  This model includes resistance contributions from the ship's hull, wave breaking, and rudder; in addition to the thrust current and swirl from the propellers.  The total axial velocity in the wake is described as
$u = u_{p}-\left(u_{f}+u_{w}+u_{r}\right)$,
where $u_{p}$ is the thrust current from the propellers, $u_{f}$ is the velocity deficit due to the frictional resistance of the hull, $u_{w}$ is the manifestation of the wave making resistance in the axial velocity profile, and $u_{r}$ is the velocity deficit due to the resistance of the twin rudders.

The ship's frictional drag current is approximated as a double Gaussian profile of the form
\begin{equation}
u_{f} = u_{Fmax} \exp\left(-3.5\tilde{a}\left[\left(\frac{y\pm0.3B}{y_h}\right)^2+\left(\frac{z}{z_h}\right)^2\right]\right),
\end{equation}
where  $u_{Fmax}$ is the maximum value of the velocity deficit due to the ship's frictional resistance at the initial data plane, B is the ship beam, $y_h$ and  $z_h$ are the half widths of the initial wake, and $\tilde{a}=2.6$ based on experimental data from \cite{pao73}.  This profile was adjusted relative to the standard Gaussian profile provided by Miner to more closely match the experimental data reported by \cite{swe87}.  

The maximum value of the frictional velocity deficit, $u_{Fmax}$, is found based on
\begin{equation}
u_{F max}=\frac{\tilde{a}U_0 S C_{f}}{\pi y_h z_h},
\end{equation}
where $C_f = \frac{0.075}{\left(\log_{10}\left(Re\right)-2\right)^2}$ is the coefficient of frictional resistance described by the ITTC-1957 correlation, $Re$ is the Reynolds number, $S$ is the wetted surface area, and $U_0$ is the ship's velocity.

Due to the presence of the free surface, the wake half widths are estimated independently.  The lateral length scale is approximated as an axisymmetric wake: 
\begin{equation}
y_h = \theta c_\delta \left[\frac{x-x_0}{\theta}\right]^\frac{1}{3}
\end{equation}
where $\theta$ is the momentum radius, $x_0$ is the virtual origin of the wake, $x$ is the location of the initial data plane, and $c_{\delta}$ is a scaling constant.  The initial data plane is taken x/L $= 0.5$ downstream of the transom to allow sufficient distance for the wake to reach nearly self--similar conditions.  

The momentum radius for an axisymmetric wake as shown by \cite{nak05} is
\begin{equation}
\theta^2=\frac{1}{8}C_D B^2
\end{equation}
where $C_D$ is the drag coefficient.  In this analysis, $C_D = C_f \frac{S}{b^2}$ is the frictional resistance non-dimensionalized by a reference area of $B^2$.
The virtual origin is then solved via
\begin{equation}
x_0/\theta=\left(\dfrac{-\sqrt{8}\alpha}{c_\delta \sqrt{C_D}}\right)^3
\end{equation}
where $\alpha= 0.5$ and $c_{\delta}= 1.14$ based on high Reynolds number data reported by \cite{joh03}.
Assuming the same virtual origin for the horizontal and vertical wakes, Miner provides the ratio of scales as
$y_h/z_h =2(B/(2D))^\frac{2}{3}$ 
where D is the ship draft.

The turbulent kinetic energy (TKE) of the ship's drag current exhibits a double peak, corresponding to the production of turbulence in the boundary layer on either side of the hull.  Miner modeled the TKE as
\begin{equation}
k_{d}= k_{max} \left(\tilde{a_1} \exp\left\{-\tilde{a_2} \left[\left(\frac{y}{y_h}\right)^2+\left(\frac{z}{z_h}\right)^2\right]\right\}-\tilde{a_3} \exp\left\{-\tilde{a_4}\left[\left(\frac{y}{y_h}\right)^2+\left(\frac{z}{z_h}\right)^2\right]\right\}\right)
\end{equation}
where $k_{max}$ is the peak value of the turbulent kinetic energy, $\tilde{a_1} = 1.35$, $\tilde{a_2} = 1.0$, $\tilde{a_3} = 0.45$, and $\tilde{a_4} = 8.22$.  Miner solves for $k_{max}$ as 
${\sqrt{k_{max}}}/{u_{Fmax}} =0.3$.
A portion of the wave--making resistance is also manifest in the velocity deficit based on formulation similar to Miner.  Here the wave--breaking contribution to the velocity profile is estimated as
\begin{equation}
u_{w} = u_{Wmax} \exp\left[-6\tilde{a}\left\{\left(\frac{z}{z_{h w}}\right)^2+\left(\frac{y\pm0.6B}{y_{h w}}\right)^2\right\}\right],
\end{equation}
where $u_{Wmax}$ is the maximum value of the velocity deficit due to wave breaking, and $y_{hw}$ and $z_{hw}$ are the half widths of the wave--breaking wake.
The maximum velocity deficit is calculated based on
\begin{equation}
u_{Wmax} =  \frac{\tilde{a} R_{wave}}{\pi \rho U_0 y_{hw} z_{hw}}
\end{equation}
where $R_{wave}$ is the wave making resistance of the hull based on the experimental data of \cite{oli01}.

The ship propellers are modeled as thrust currents with swirl with an axial velocity profile given as
\begin{equation}
u_{p}=u_{pMax} exp\left[-2\tilde{a}\left(\frac{r}{r_h}\right)^2\right]
\end{equation}
where  $\tilde{a}= 2.45$, the constant 2 is a profile tuning value, and $r_h=0.086x$.  The maximum axial velocity is given by
\begin{equation}
u_{pMax}= \frac{\tilde{a} T}{\pi \rho U_0 (r_h)^2}
\end{equation}
where $T$ is the total thrust of the propellers to overcome the combined frictional and wave--making resistance.
The swirl velocity is based on the self-similar propeller velocity.
\begin{equation}
V_s=V_{sMax} \tilde{a_1} \left(\frac{r}{r_{sh}}\right) \exp \left[-\tilde{a_2}\left(\frac{r}{r_{sh}}\right)^2\right]
\end{equation}
where $\tilde{a_1}=4.34$, $\tilde{a_2}=3.56$, and $r_{sh}$ is half width of the swirl profile taken to be
$r_{s h}=0.095 \sqrt{xB}$.
The maximum swirl velocity, $V_{s Max}$, is a function of the propeller torque ($Q$)
\begin{equation}
V_{sMax}= \frac{Q}{\pi \rho U_0 \frac{\tilde{a_1}}{\tilde{a_2}}(r_{s h})^3 }.
\end{equation}
The propeller torque is a function of the swirl number, $\tilde{S}$, and the propeller diameter, $D_p$
\begin{equation}
Q = \tilde{S} \left(T\frac{D_p}{2}\right),
\end{equation}
where the swirl number is taken to be $S$ = 0.3.

The self--similar turbulent kinetic energy from the propeller wake is given by
\begin{equation}
k_p=k_{p Max} \exp\left[-\tilde{a}\left(\frac{r}{r_h}\right)^2\right],
\end{equation}
where  $k_{pMax}= (0.59 u_{Fmax})^2$.

The rudders are modeled as NACA 0015, with $C_{d_0}=0.0093$.  Single Gaussian profiles are used to model the wake deficits as outlined by Miner.
\begin{equation}
u_{r} = u_{rMax} exp\left[-\tilde{a}\left\{\left(\frac{y \pm y_{r}}{y_{hr}}\right)^2+\left(\frac{z - z_{r}}{z_{hr}}\right)^2\right\}\right]
\end{equation}
where $y_r$ and $z_r$ are the rudder horizontal and vertical offsets from the surface centerline.  The maximum velocity deficit due to the rudder is estimated as
\begin{equation}
u_{rMax} = \frac{\tilde{a_r} R_{rudder}}{\pi \rho U_0 y_{hr} z_{hr}}
\end{equation}
where $R_{r}$ is the resistance of the rudder and $y_{hr}$ and $z_{hr}$ are the wake half widths.  The vertical half width, $z_{hr}=\frac{1}{2} b_{r}$, where $b_{r}$ is the rudder span, while the horizontal half width, $y_{hr}$ is set as
\begin{equation}
y_{hr} = \tilde{a}_r\sqrt{x t_{r}}+\frac{1}{2}t_{r};
\end{equation}
where $t_{r}$ is the thickness of the rudder and $\tilde{a}_r = 0.23 \sqrt{C_{dr}}$.
The TKE due to the rudders is found by using equation (A6) with
$\sqrt{k_{max}}/u_{rMax} =1$.
%

Bilge keels are not included in the current model as their contribution to the velocity profile is considered negligible.  Miner provides a formulation for the inclusion of bilge vortices, but the experimental data presented by \cite{swe87} and \cite{hoe97} suggest that the pair of bilge vortices are ingested by the propellers and their rotation canceled.  

The total turbulent kinetic energy is taken to be the sum of the contributions from ship drag current and the propeller
\begin{equation}
k=k_{d}+k_{p}.
\end{equation}
The turbulent dissipation rate of the drag current $\varepsilon_{d}$ and the thrust current $\varepsilon_{p}$ are modeled based on \cite{has80} as:
\begin{equation}
\varepsilon_{d}= \sqrt{\frac{3}{A}} \left(k_{d}\right)^\frac{3}{2}, \varepsilon_{p}=  \frac{\sqrt{12}}{B} \left(k_{p}\right)^\frac{3}{2}
\end{equation}
where 
\begin{equation}
A=C_d B^2  \frac{U_0}{8 u_{fMax}}.
\end{equation}

\end{document}